\documentclass[jmp,reprint,sd,amsmath,amssymb,graphicx]{revtex4-1}

\draft 
\usepackage{graphicx}
\usepackage{dcolumn}
\usepackage{bm}
\begin{document}

\preprint{AIP/123-QED}


\title[Electrodynamics in six dimensions]{Electrodynamics in flat spacetime
of six dimensions} 



\author{Yurij Yaremko}
\email[Electronic mail: ]{yar@icmp.lviv.ua}
\affiliation{Institute for Condensed Matter Physics\\ Svientsitskii Str. 1, 79011 Lviv, Ukraine}


\date{\today}

\begin{abstract}
We consider the dynamics of a classical charge in flat spacetime of six dimensions. The mass shell relation of a free charge admits nonlinear oscillations. Having analyzed the problem of on eigenvalues and eigenvectors of Faraday tensor, we establish the algebraic structure of electromagnetic field in 6D. We elaborate the classification scheme based on three field's invariants. Using the basic algebraic properties of the electromagnetic field tensor we analyze the motion of a charge in constant electromagnetic field. Its world line is a combination of hyperbolic and circular orbits which lie in three mutually orthogonal sheets of two dimensions. Within the braneworld scenario, we project the theory on the de Sitter space of four dimensions. Actually, as it turns out, spins of elementary particles themselves are manifestations of extra dimensions.
\end{abstract}



\maketitle 


\section{Introduction}\label{Intro}

In the string theory, extra dimensions are required to ensure the mathematical consistency of the theory \cite{S-TYN2010}. They are typically thought to be very small, close to the Planck length. Dimensions higher than four arise in braneworld scenarios where the observable $(3+1)$ Universe is embedded in a higher dimensional bulk (see \cite{CBR2016} and references therein). The extra dimensions can be very large or even infinitely extended. Search of manifestations of extra dimensions (e.g., tiny black holes \cite{NA-HDD1998,AFK2015}) continues at the Large Hadron Collider \cite{ATLAS2016}. It is likely that a higher energy is needed for the experiments to be successful.

Thus. the question arises what are possible manifestations of extra dimensions when we put an electromagnetic interaction into a spacetime of higher dimension appeared in braneworld scenarios. In the modern language of differential geometry the Maxwell equations in Minkowski space of $D$ dimensions look as follows
\begin{equation}\label{Max-D}
{\rm d}\hat{F}=0,\qquad {\rm d}^{\,\star}\!{\hat F}=\Omega_{D-1}\!\!\!\phantom{1}^{\,*}{\hat j},
\end{equation}
where $\Omega_{D-1}$ is the area of unit $(D-1)$-dimensional sphere. At each point of Minkowski space ${\mathbb M}_{\,D}$, a given current density $\hat{j}$ generates the electromagnetic field which is determined by the Faraday $2$-form $\hat F$ and Hodge dual $(D-2)$-form $\!\!\!\phantom{1}^{\,\star}\!{\hat F}$. As the the Faraday 2-form is exact ${\hat F}={\rm d}{\hat A}$, the equations can be easily transformed into wave equation $\Box A_\mu=-\Omega_{D-1}j_\mu$ on the components of the $1$-form potential ${\hat A}$. Symbol $\Box$ denotes d'Alembert differential operator. In curved spacetime the problem is analyzed in Ref.~\cite{C-BCL2006}.

For a point-like source the wave equation has been solved with the help of either Green's function method \cite{IvSk1951,TY2012} or elegant iterative procedure \cite{KosIJMP2008}. The solutions extremely grow at the immediate vicinity of the point where the charge is placed. To obtain sensible equation of motion of a point-like charge under the influence of an external electromagnetic field as well as its own field, the regularization procedure is necessary. In Ref.~\cite{KosTMP99} the derivation of analogue of the Lorentz-Abraham-Dirac equation in six dimensions is patterned after Dirac]s seminal paper \cite{Dirac1938}. Kosyakov produced a proper relativistic equation of motion of radiating charge via analysis of energy-momentum conservation in $6D$. Alternatively, the method of removing inevitable infinities from Green's function which uses functional analysis tools is developed in Ref.~\cite{KLSprd02} for electrodynamics in an arbitrary dimensions.

Consistent elimination procedure can be based on the conserved quantities arising from the symmetry of the problem. Minkowski spacetime isometries constitute the group of symmetry of the electrodynamics in $D$ dimensions. The Poincar\'e group denoted as  ${\cal P}(1,D-1)$ is a semidirect product of the Abelian group of space and time translations ${\mathbb R}^{1,D-1}$ and generalized orthogonal group, or Lorentz group $O(1,D-1)$. Generally, the regularization procedure which exploits the symmetry properties of the theory manipulates with $D(D+1)/2$ integrals of motions corresponding to transformations from ${\cal P}(1,D-1)$. Careful analysis of energy-momentum and angular momentum balance equations yields the renormalized equations of motion of radiating charge in six dimensions \cite{YarJPA04}. In Ref.~\cite{MMtmp2008} the expressions for radiated energy-momentum have been obtained for an arbitrary even dimensions.

Similar to gravity \cite{NA-HDD1998}, the Li\`{e}nard-Wiechert potential in six dimensions is stronger than that in four dimensions. There are three types of divergences in $6D$. Their elimination is not a trivial matter. Conservation laws are an immovable fulcrum about which tips the balance of truth regarding renormalization, radiation reaction, and particle's individual characteristics. It turned out that in six dimensions, for a given $\delta$-like current density $\hat{j}$, two renormalization parameters are necessary to absorb inevitable divergencies coming from particle's self-action. Apart from usual ``bare mass'' $m_0$, the particle action term involves an additional renormalization constant, say $\mu_0$, which absorbs one extra divergent term. The  particle's dynamics is governed by the action
\begin{equation}\label{Spart}
S_{\rm part}=\int{\rm d}\lambda \gamma^{-1}(\dot{z})\left(-m_0-\frac12\mu_0 K(\dot{z},\ddot{z})\right),
\end{equation}
where $\gamma=1/\sqrt{-\dot{z}^2}$ is the Lorentz factor. The Lagrangian depends on the squared curvature of the particle's world line
\begin{equation}\label{k2}
K(\dot{z},\ddot{z})=\gamma^4\left(\ddot{z}^2+\gamma^2(\dot{z}\cdot\ddot{z})^2\right).
\end{equation}
Six functions $z^\alpha(\lambda)$ where index $\alpha$ runs from $0$ to $5$ parameterize the points $z(\lambda)$ of particle's world line $\zeta:{\mathbb R}\to{\mathbb M}_{\,6}$. The derivatives $\dot{z}^\alpha(\lambda)={\rm d}z^\alpha(\lambda)/{\rm d}\lambda$ are the components of a tangent vector for $\zeta$ at the point $z(\lambda)\in\zeta$. Two points over $z^\alpha(\lambda)$ denotes the second derivatives $\ddot{z}^\alpha(\lambda)={\rm d}^2z^\alpha(\lambda)/{\rm d}\lambda^2$.

It is worth noting that the action (\ref{Spart}) governs the dynamics of the {\it rigid particle} \cite{PavAACA2015}. It is one-dimensional approximation \cite{PavPLB1988,GILNPLB1989} of Polyakov-Kleinert string whose action integral contains, apart from Nambu-Goto term, an additional term depending on the curvature of the world sheet of the string \cite{PolNPB1986,KlnPLB1986,PavAACA2015}. A classical point charge in renormalizable $6D$ electrodynamics is not structureless.

Our concern in this paper is with the electrodynamics in six dimensions. We consider the behaviour of a {\it test charge}, i.e. a point charged particle which itself does not influence the field. In Section \ref{FreeRigid} we study inertial properties of a free charge in six dimensions. They are highly non-trivial because the Lagrangian for the point-like analogue of the Polyakov-Kleinert string depends on acceleration. In Section \ref{Field-6D} we find particular solutions of the Lorentz force equation in $6D$. In Section \ref{Algebr_6D} the algebraic structure of electromagnetic field in six dimensions is investigated. We propose classification scheme based on three field's invariants. In Section \ref{Compact6-4} we reduce two extra dimensions by means of specific surjective diffeomorphism that projects flat spacetime ${\mathbb M}_{\,6}$ onto four-dimensional se Sitter space.  In Section \ref{Concl}, we summarize the main ideas and results.


\section{Free rigid particle}\label{FreeRigid}

The aim of this Section is to explain the mechanical properties of an elementary charge in six dimensions. The obvious difference from elementary charged particles in four dimensions is that they have at their disposal two extra dimensions. It is important, but not principled. The crucial difference between the two objects is that the action (\ref{Spart}) contains higher derivatives. For this reason the relation between energy and momentum of a free rigid particle has highly non-trivial form. Before going to the mass shell relation we consider the equations of motion.

\subsection{Hamiltonian equations of motion}\label{Hamilton}

For reader's convenience we recall briefly the dynamics governed by the particle action (\ref{Spart}). We follow the analysis presented in Ref.~\cite{PlushPLB91} with the only difference being that the particle moves in flat spacetime of six dimensions (see also \cite[\S 10.1]{KosBook07}). We also use the results presented in \cite{ASWactaB1989,DHOTPLB1990}.

Putting the components $\dot{z}^\mu$ of particle's six-velocity as new coordinates $q^\mu$ we pass to the Lagrangian which depends on the first-order derivatives: 
\begin{equation}\label{Ltilde}
\tilde{L}=-\sqrt{-q^2}\left(m+\frac12\mu
K(q,\dot{q})\right)-\Theta^\mu\left(\dot{z}_\mu-q_\mu\right),
\end{equation}
where $\Theta^\mu$ are Lagrange multipliers. We assume that both already renormalized constants, $m$ and $\mu$, are finite and observable parameters. Let us define the Hamiltonian function.

Defining the canonical momenta $P_\mu=\partial\tilde{L}/\partial\dot{z}^\mu$ and $p_\mu^\Theta=\partial\tilde{L}/\partial\dot{\Theta}^\mu$, we obtain two primary constraints $P_\mu+\Theta_\mu\approx 0$ and
$p_\mu^\Theta\approx 0$. They have a non-zero Poisson bracket. Passing to the Dirac's brackets, we exclude the pair $(\Theta^\mu,p_\mu^\Theta)$ of canonically conjugated variables.

For future convenience we introduce six-vector $X^\mu_\bot=X^\mu+\gamma^2(q)(X\cdot q)q^\mu$. As the scalar product $(q\cdot q)=-\gamma^{-2}(q)$, the vector is orthogonal to six-velocity, namely $(X_\bot\cdot q)=0$. Differentiating the Lagrangian (\ref{Ltilde}) with respect to $\dot{q}^\mu$ we obtain the six-momentum canonically conjugated to $q^\mu$:
\begin{equation}\label{pi-mu}
\pi_\mu=-\mu\gamma^3(q)\dot{q}_{\bot,\mu}.
\end{equation}
There exists the constraint
\begin{equation}\label{constr1}
\phi_1=(\pi\cdot q)\approx 0.
\end{equation}

Having performed the Legendre transform we derive the Hamiltonian function from the Lagrangian (\ref{Ltilde}). Using the expression $\pi^2=\mu^2\gamma^2(q)K(q,\dot{q})$ we construct canonical Hamiltonian
\begin{eqnarray}\label{Hc}
H&=&\dot{z}^\mu P_\mu+\dot{q}^\mu\pi_\mu+\dot{\lambda}^\mu
p_\mu^\lambda -\tilde{L}\nonumber\\
&=&P\cdot q+\gamma^{-1}\left(m+\frac{q^2\pi^2}{2\mu}\right).
\end{eqnarray}
There is the secondary constraint $\phi_2=H\approx 0$ originating from the time derivative $\dot{\phi}_1\approx 0$. As the Poisson bracket $\left\{\phi_1,\phi_2\right\}=-\phi_2$, other constraints could not be found.

The total Hamiltonian $H'=H+\upsilon_1\phi_1$ produces the following equations of motion
\begin{eqnarray}
\dot{z}_\mu&=&q_\mu,\label{HamEqs1}\\[1ex]
\dot{P}_\mu&=&0,\label{HamEqs2}\\[0.5ex]
\dot{q}_\mu&=&-\gamma^{-3}(q)\frac{\pi_\mu}{\mu}+\upsilon_1q_\mu,\label{HamEqs3}\\
\dot{\pi}_\mu&=&-P_\mu+m\gamma(q)q_\mu-\frac{3\pi^2}{2\mu}\gamma^{-1}(q)q_\mu-\upsilon_1\pi_\mu.\label{HamEqs4}
\end{eqnarray}
Comparing the equation (\ref{HamEqs3}) with the momentum defined by eq.~(\ref{pi-mu}) we obtain the multiplier
\begin{equation}\label{v1}
\upsilon_1=-\gamma^2(q)(\dot{q}\cdot q).
\end{equation}
Inserting this into expression (\ref{HamEqs4}) and substituting the time derivative of the right hand side of eq.~(\ref{pi-mu}) for $\dot{\pi}_\mu$ we derive the total particle's six-momentum $P$ as function of $q$, $\dot{q}$, and $\ddot{q}$:
\begin{align}\label{P-HamEqs4}
P_\mu&=\gamma\left(m-\frac12\mu K(q,\dot{q})\right)q_\mu+3\mu\gamma^5(q\cdot\dot{q})\dot{q}_{\bot,\mu}\nonumber\\
&+\mu\gamma^3\ddot{q}_{\bot,\mu}.
\end{align}

We restrict ourselves to the sector of time-like world lines $\dot{z}^2<0$, $\ddot{z}^2_\bot\geq 0$. Since $K=\gamma^4\ddot{z}^2_\bot$, the squared curvature (\ref{k2}) is positively defined here. Passing to the parametrization by the proper time $\tau$, we express the conserved momentum in terms of normalized six-velocity $u^\nu\equiv{\rm d}z^\nu/{\rm d}\tau=\gamma\dot{z}^\nu$, six-acceleration $a^\nu\equiv{\rm d}u^\nu/{\rm d}\tau=\gamma^2\ddot{z}^\nu_\bot$, and its proper time-derivative $\dot{a}^\nu\equiv{\rm d}a^\nu/{\rm d}\tau=\gamma^3\left[\dddot{z}^\nu_\bot+3\gamma^2\left(\ddot{z}\cdot\dot{z}\right)\ddot{z}^\nu_\bot
+\gamma^2\ddot{z}^2_\bot\dot{z}^\nu\right]$:
\begin{equation}\label{P}
P^\nu=mu^\nu+\mu\left(\dot{a}^\nu-\frac32(a\cdot a)u^\nu\right).
\end{equation}
By the normalized six-velocity $u(\tau)$ we mean the time-like  tangent vector with norm $1$: $(u\cdot u)\equiv\eta_{\alpha\beta}u^\alpha u^\beta=-1$. In this parametrization the squared curvature $K=(a\cdot a)$.

\subsection{Energy, momentum and mass shell}

Adapting \cite[eq.(12)]{ASWactaB1989}, we express the mass shell relation in the form of the second order differential equation on the squared curvature
\begin{equation}\label{MShell}
-(P\cdot P)=\left(m-\frac32\mu K\right)\left(m-\frac12\mu K\right)-\frac12\mu^2\ddot{K}.
\end{equation}
The other important relation is
\begin{equation}\label{P_u}
(P\cdot u)=-m+\frac12\mu K.
\end{equation}
The particular solution $K=0$ yields the standard mass shell of structureless particle: $-(P\cdot P)=m^2$ where six-momentum $P$ and six-velocity $u$ are collinear: $P^\mu=mu^\mu$. Further we are interesting in the mass shell of rigid charged particle with positively defined squared curvature. In this case the momentum of charge and its six-velocity are not collinear. It depends on the particle's acceleration and its time derivative (see eq.~(\ref{P})).

The equation (\ref{MShell}) defines the non-linear oscillations parameterized by Jacobi elliptic functions \cite{DHOTPLB1990}. We overmultiply it on $\dot{K}$ and then integrate both the left side and the right side of this equation. We express the result as the sum of kinetic energy $\mu^2\dot{K}^2/2$ and potential $W(K)=-\mu^2K^3/2+2m\mu K^2-2(m^2+P^2)K$. Since the vibration of free rigid charge has no friction nor forcing, the energy (constant of integration) is conserved quantity: $\mu^2\dot{K}^2/2+W(K)=E$.

\begin{figure}[ht]
\begin{center}
\includegraphics*[scale=0.8,angle=0,trim=4 4 0 4]{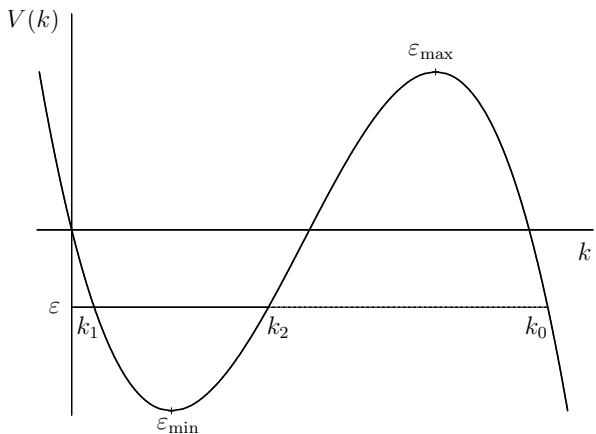}
\end{center}
\caption{Graph of scaled potential $V(k)=-k^3+2k^2-(1-\delta)k$ with parameter $0<\delta<1$. Energy changes from $\varepsilon_{\rm min}$ (minimum) to $\varepsilon_{\rm max}$ (maximum). If $\varepsilon=\varepsilon_{\rm min}$ the squared curvature is constant: $k=k_1(\delta,\varepsilon_{\rm min})=k_2(\delta,\varepsilon_{\rm min})$. For a given value of energy $\varepsilon_{\rm min}<\varepsilon\leq 0$ the scaled curvature $k$ oscillates between roots $k_1$ and $k_2$. Together with zeroth root, $k_0$, they define the period of oscillation and its frequency.
}\label{V_K}
\end{figure}

To simplify further consideration we introduce the dimensionless variables
\begin{equation}\label{dimless}
\tau'=\sqrt{\frac{2m}{\mu}}\,\tau,\quad k=\frac{\mu}{2m}K,
\end{equation}
and ruling parameters:
\begin{equation}\label{rulpar}
\delta=\frac{-(P\cdot P)}{m^2},\quad \varepsilon=\frac{\mu E}{4m^3}.
\end{equation}
We obtain the first order differential equation
\begin{equation}\label{K-mshell}
\left(\frac{{\rm d}k}{{\rm d}\tau'}\right)^2=k^3-2k^2+(1-\delta)k+\varepsilon,
\end{equation}
which simply expresses the fact that the total energy is the sum of kinetic and potential energies. The roots of the cubic polynomial in the right hand side define the solution to this equation. The oscillations take place if and only if all the roots are distinct and real. They can be expressed in terms of trigonometric functions
\begin{equation}\label{K-roots}
k_j(\delta,\varepsilon)=\frac{2}{3}\left[1+\sqrt{1+3\delta}\cos\left(\psi(\delta,\varepsilon)+\frac{2\pi}{3}j\right)\right],
\end{equation}
where index $j$ runs from $0$ to $2$ and the phase is
\begin{equation}\label{psi-roots}
\psi(\delta,\varepsilon)=
\frac{1}{3}\arccos\left(\frac{-1+9\delta-27\varepsilon/2}{\left(1+3\delta\right)^{3/2}}\right).
\end{equation}
Energy parameter changes from minimum
\begin{equation}\label{eps-min}
\varepsilon_{\rm min}(\delta)=-\frac{2}{27}\left[1-9\delta+(1+3\delta)^{3/2}\right],
\end{equation}
where phase $\psi=0$, to maximum
\begin{equation}\label{eps-max}
\varepsilon_{\rm max}(\delta)=\frac{2}{27}\left[-1+9\delta+(1+3\delta)^{3/2}\right],
\end{equation}
where $\psi=\pi/3$. They are extrema points of potential function $V(k)$ which is pictured in Fig.~\ref{V_K}.

Factoring the cubic polynomial in the right hand side of eq.~(\ref{K-mshell}) we rewrite it as follows
\begin{equation}\label{K-120}
\left(\frac{{\rm d}k}{{\rm d}\tau'}\right)^2=\left(k-k_1\right)\left(k-k_2\right)\left(k-k_0\right).
\end{equation}
According to the Handbook \cite[17.4.61]{AbrStg}, the solution is
\begin{equation}\label{K-12}
k=k_1+\left(k_2-k_1\right){\rm sn}^2(w\backslash\alpha),
\end{equation}
where the argument of the elliptic sine is
\begin{equation}\label{w-4-62less}
w=\frac12\sqrt{k_0-k_1}\,\tau'+\phi_0,
\end{equation}
and the modular angle
\begin{equation}\label{alph-4-62less}
\sin^2\alpha=\frac{k_2-k_1}{k_0-k_1}.
\end{equation}
The phase $\phi_0$ is given by initial conditions.
The solution is presented in Ref.~\cite[eq.~(17)]{DHOTPLB1990}.

According to eq.~(\ref{K-12}), for a given parameter $\varepsilon$
the variable $k$ ``moves'' forward and backward from $k_1$ to $k_2$
(see Fig.~\ref{V_K}). If $\varepsilon$ is greater than $0$, the orbit
contains the segment where $k<0$. Such a trajectory is non-physical.

The form of a curvature orbit heavily depends on the values of renormalization constant
and parameter $\delta$. If $\mu>0$ and $\delta>1$, the scaled potential
$V(k)$ does not admit a bounded orbit without a ``negative'' segment.
In this case positively defined solution \cite[eq.~(16)]{DHOTPLB1990} contains the term
which is inversely proportional to the squared elliptic sine; its argument and
modular angle are given by eqs.~(\ref{w-4-62less}) and (\ref{alph-4-62less}),
respectively. The solution describes non-linear oscillations when the curvature goes
to infinity over a period. 

\subsection{The rest frame of a free rigid particle}\label{RestFrame}

In this Paragraph we solve the system of non-linear differential equations (\ref{P}) where the squared curvature $(a\cdot a)$ is given by eq.~(\ref{K-12}). To simplify the equations as much as possible we pass to the coordinate system where the translational momentum is $\mathbf{P}=0$. In this privileged reference frame the expression (\ref{P_u}) depends on the zeroth component of the normalized six-velocity only. Solving it we obtain
\begin{align}\label{P-u_own}
u^0(\tau')&=\frac{1-k(\tau')}{\sqrt{\delta}},\nonumber\\
&=\frac{1}{\sqrt{\delta}}\left[1-k_1-\left(k_2-k_1\right){\rm sn}^2(w\backslash\alpha)\right]
\end{align}
where $\delta=(P^0/m)^2$. Direct calculations show, that the function satisfies the equation of motion
\begin{equation}\label{u_0own}
\frac{{\rm d}^2u^0}{{\rm d}\tau'^2}+\frac{1-3k(\tau')}{2}u^0=\frac{{\sqrt{\delta}}}{2},
\end{equation}
which is the zeroth part of eq.~(\ref{P}) in terms of dimensionless variables.

To find out the zeroth coordinate function we integrate the zeroth component over the proper time variable. Using the definition \cite[16.25.1]{AbrStg} we obtain
\begin{equation}\label{z0-4-69}
z^0(\tau')=z^0_0+\frac{1-k_1}{\sqrt{\delta}}\tau'-\frac{2(k_2-k_1)}{\sqrt{\delta(k_0-k_1)}}{\rm Sn}(w\backslash\alpha).
\end{equation}
Using the relation \cite[16.26.1]{AbrStg} one can substitute the elliptic integral of the second kind ${\rm E}(w\backslash\alpha)$ for the function ${\rm Sn}(w\backslash\alpha)$. 

\begin{figure}[ht]
\begin{center}
\includegraphics*[scale=0.9,trim=5 0 4 4, angle=0]{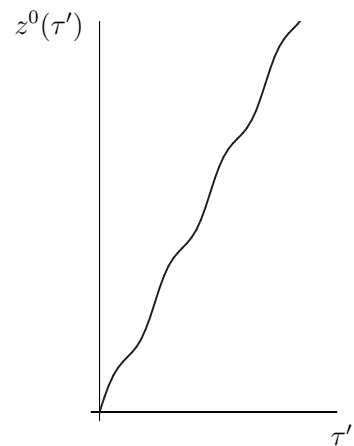}
\end{center}
\caption{Depiction of the evolution of time coordinate of free rigid particle. Space coordinates perform periodic non-linear oscillations parameterized by Jacobi elliptic functions.
}\label{z0-001}
\end{figure}

At the minimum of the cubic potential pictured in Fig.~\ref{V_K} the roots $k_2=k_1=k_{\rm min}$ and, therefore, the squared curvature given by eq.~(\ref{K-12}) does not change with time. Inserting $k_{\rm min}=(2-\sqrt{1+3\delta})/3$ in eq.~(\ref{P-u_own}) we obtain
\begin{equation}\label{u0-17.4.61own}
u^0_{\rm min}=\frac{1+\sqrt{1+3\delta}}{3\sqrt{\delta}}.
\end{equation}
Rigid particle moves uniformly in time coordinate.

In terms of dimensionless variables the space part of eq.~(\ref{P}) looks as follows:
\begin{equation}\label{u_iown}
\frac{{\rm d}^2u^i}{{\rm d}\tau'^2}+\frac{1-3k(\tau')}{2}u^i=0,
\end{equation}
If $\varepsilon=\varepsilon_{\rm min}$, $k(\tau')=k_{\rm min}$ and we deal the second order differential equation on the space components of six-velocity. The solution is
\begin{equation}\label{u-17.4.61own}
\mathbf{u}_{\rm min}=\mathbf{A}\cos\left(\omega\tau\right)+\mathbf{B}\sin\left(\omega\tau\right),
\end{equation}
where the squared frequency
\begin{equation}\label{omega-17.4.61}
\omega^2=\frac12\left(\sqrt{1+3\delta}-1\right).
\end{equation}
By symbols $\mathbf{A}$ and $\mathbf{B}$ we denote the orthogonal five-vectors of equal magnitudes $|\mathbf{A}|=|\mathbf{B}|=|\mathbf{u}_{\rm min}|$. The magnitude of spatial five-vector of six-velocity is simply $|\mathbf{u}|=\sqrt{(u^0-1)(u^0+1)}$. Inserting eq.~(\ref{u0-17.4.61own}) we obtain
\begin{equation}\label{AB-17.4.61}
|\mathbf{u}_{\rm min}|=\sqrt{\frac{2\left(2-\sqrt{1+3\delta}\right)}{3\left(\sqrt{1+3\delta}-1\right)}}.
\end{equation}
Recall that the parameter $0<\delta<1$. At the minimum of curvature the rigid particle moves along the helical line
\cite{ASWactaB1989,DHOTPLB1990}.

Near the minimum of potential pictured in Fig.~\ref{V_K} where $\varepsilon>\varepsilon_{\rm min}$ but $k_2-k_1\ll 1$, the elliptic sine in the expression for squared curvature (\ref{K-12}) can be approximated by trigonometric sine. Putting this in equation of motion (\ref{u_iown}) we obtain the Mathieu equation \cite[S.~20]{AbrStg}. We do not bother with it. Instead, we present the exact solution of eq.~(\ref{u_iown}) for zeroth energy level $\varepsilon=0$. In this specific case the roots $k_1=0$, $k_2=1-\sqrt{\delta}$, and $k_0=1+\sqrt{\delta}$. The zeroth component of six-velocity takes the form
\begin{equation}\label{u0-4-62less}
u^0(\tau')=\frac{1}{\sqrt{\delta}}\left[1-\left(1-\sqrt{\delta}\right){\rm sn}^2(w\backslash\alpha)\right],
\end{equation}
where the argument $w=\frac12\sqrt{1+\sqrt{\delta}}\,\tau'+\phi_0$ and the modular angle $\sin^2\alpha=(1-\sqrt{\delta})/(1+\sqrt{\delta})$. The magnitude of five-vector $\mathbf{u}$ depends on elliptic functions:
\begin{equation}
|\mathbf{u}(\tau')|=\sqrt{\frac{1-\delta}{\delta}}{\rm cn}(w\backslash\alpha){\rm dn}(w\backslash\alpha).
\end{equation}
We assume $\mathbf{u}(\tau')=\mathbf{C}\,{\rm cn}(w\backslash\alpha){\rm dn}(w\backslash\alpha)$ where $\mathbf{C}$ is an arbitrary constant five-vector with length $|\mathbf{C}|=\sqrt{(1-\delta)/\delta}$.
Having integrated this function over time we obtain coordinate five-vector
\begin{equation}\label{z1-4-69more}
\mathbf{z}(\tau')=\mathbf{z}_0+\frac{2\mathbf{C}}{\sqrt{1+\sqrt{\delta}}}{\rm sn}(w\backslash\alpha).
\end{equation}
The solution describes the periodic orbit, repeating itself in a sinusoidal fashion with constant amplitude and constant frequency. 

\section{Rigid charge in electromagnetic field}\label{Field-6D}

Following \cite{KosTMP99,KLSprd02}, we introduce minimal coupling between rigid charge $e$ and an external electromagnetic field. We add the interaction term
\begin{equation}\label{Sint}
S_{\rm int}=e\int{\rm d}\lambda A_\mu\dot{z}^\mu,
\end{equation}
where $A_\mu(z)$ are the components of the electromagnetic one-form potential $\hat{A}=A_\mu{\rm d}x^\mu$ evaluated at point $z(\lambda)$ on the world line $\zeta$ where the charge $e$ is placed. Variation of the total action $S=S_{\rm part}+S_{\rm int}$ yields the equations of motion ${\rm d}P_\mu/{\rm d}\lambda=eF_{\mu\nu}q^\nu$ where the momentum $P_\mu$ is given by eq.~(\ref{P-HamEqs4}). Passing to the proper time parametrization we obtain the analogue of the Lorentz-force equation of motion in six dimensions:
\begin{equation}\label{Lf-6D}
\frac{{\rm d}}{{\rm d}\tau}\left[mu_\mu+\mu\left(-\ddot{u}_\mu+\frac32(a\cdot a)u_\mu\right)\right]=eF_{\mu\nu}u^\nu.
\end{equation}
Here $e$ is the magnitude of electric charge and $F_{\mu\nu}=\partial_\nu A_\mu-\partial_\mu A_\nu$ are the components of the electromagnetic field 2-form $\hat{F}={\rm d}\hat{A}$. If the Minkowski rectangular coordinates are adapted, the state of electromagnetic field at point $x\in{\mathbb M}_{\,6}$ is specified by five components $(E^1,E^2,E^3,E^4,E^5)$ of electric field five-vector and ten elements $F_{ik}$, $i,k=\overline{1,5}$, $i<k$ of the skew symmetric matrix:
\begin{equation}\label{Far}
(F_{\alpha\beta})=\left(
\begin{array}{cccccc}
0&-E^1&-E^2&-E^3&-E^4&-E^5\\
E^1&0&F_{12}&-F_{31}&F_{14}&-F_{51}\\
E^2&-F_{12}&0&F_{23}&-F_{42}&F_{25}\\
E^3&F_{31}&-F_{23}&0&F_{34}&-F_{53}\\
E^4&-F_{14}&F_{42}&-F_{34}&0&F_{45}\\
E^5&F_{51}&-F_{25}&F_{53}&-F_{45}&0
\end{array}
\right).
\end{equation}
We rise index $\mu$ in equation (\ref{Lf-6D}) by means of the metric tensor $\eta={\rm diag}(-1,1,1,1,1,1)$:
\begin{equation}\label{Lf_6D}
\frac{{\rm d}}{{\rm d}\tau}\left[mu^\mu+\mu\left(-\ddot{u}^\mu+\frac32(a\cdot a)u^\mu\right)\right]=eF^\mu{}_\nu u^\nu.
\end{equation}
Our task is to solve the equations of motion in the specific case of static and spatially uniform electromagnetic field.

First, we suppose $\mu\ll 1$ such that the nonlinear part of equation is much smaller than the linear one. Having applied the method of successive approximation we substitute the contraction of electromagnetic field tensor and particle's six-velocity for the particle's six-acceleration:
\begin{equation}\label{a_Fu}
a^\mu=\frac{e}{m}F^\mu{}_\nu u^\nu.
\end{equation}
We replace the kinematic variables un the non-linear terms by differential consequences of this expression:
\begin{align}\label{adot_FFu}
\dot{a}^\mu&=\frac{e}{m}F^\mu{}_\nu a^\nu\nonumber\\
&=\left(\frac{e}{m}\right)^2F^\mu{}_\nu F^\nu{}_\alpha u^\alpha;\\
\ddot{a}^\mu&=\left(\frac{e}{m}\right)^3F^\mu{}_\nu F^\nu{}_\alpha F^\alpha{}_\beta u^\beta.\nonumber
\end{align}
For the squared curvature and its time derivative we obtain
\begin{align}\label{aa_FFu}
(a\cdot a)&=\left(\frac{e}{m}\right)^2\eta_{\alpha\beta}F^\alpha{}_\mu F^\beta{}_\nu u^\mu u^\nu;\\
\frac{{\rm d}(a\cdot a)}{{\rm d}\tau}&=\left(\frac{e}{m}\right)^3\eta_{\alpha\beta}F^\alpha{}_\kappa F^\kappa{}_\delta F^\beta{}_\nu u^\delta u^\nu.\nonumber
\end{align}
We take into account that tensor $\hat{F}$ is supposed to be constant.

The expressions look terribly horrible, but the matrix
\begin{equation}\label{F-ar}
(F^\mu{}_\nu)=\left(
\begin{array}{cccccc}
0&E^1&E^2&E^3&E^4&E^5\\
E^1&0&F_{12}&-F_{31}&F_{14}&-F_{51}\\
E^2&-F_{12}&0&F_{23}&-F_{42}&F_{25}\\
E^3&F_{31}&-F_{23}&0&F_{34}&-F_{53}\\
E^4&-F_{14}&F_{42}&-F_{34}&0&F_{45}\\
E^5&F_{51}&-F_{25}&F_{53}&-F_{45}&0
\end{array}
\right)
\end{equation}
describes not only the electromagnetic field. It bears the imprint of the inertial frame which is used to determine the components of electromagnetic field tensor. To solve the non-linear differential equation (\ref{Lf_6D}) we apply the technique of projection operators \cite{FradJPA78,YarJMP13} based on the eigenvectors and eigenvalues of the electromagnetic field tensor (\ref{F-ar}). Eigenvectors constitute the basis defining the linear transformation which makes this tensor diagonal. The transformation can be easily transformed into Lorentz transformation which simplifies the equations of motion substantially.

Looking ahead, there are opportunities to simplify some specific field tensors to the form
\begin{equation}\label{FL-b}
(F^\alpha{}_\beta)'=\left(
\begin{array}{cccccc}
0&b&0&0&0&0\\
b&0&0&0&0&0\\
0&0&0&0&0&0\\
0&0&0&0&0&0\\
0&0&0&0&0&0\\
0&0&0&0&0&0
\end{array}
\right),
\end{equation}
or
\begin{equation}\label{FL-a}
(F^\alpha{}_\beta)'=\left(
\begin{array}{cccccc}
0&0&0&0&0&0\\
0&0&0&0&0&0\\
0&0&0&a&0&0\\
0&0&-a&0&0&0\\
0&0&0&0&0&0\\
0&0&0&0&0&0
\end{array}
\right),
\end{equation}
etc. The most complicated tensor has the form
\begin{equation}\label{FL}
(F^\alpha{}_\beta)'=\left(
\begin{array}{cccccc}
0&b&0&0&0&0\\
b&0&0&0&0&0\\
0&0&0&a&0&0\\
0&0&-a&0&0&0\\
0&0&0&0&0&c\\
0&0&0&0&-c&0
\end{array}
\right).
\end{equation}
Field strengths $a$, $b$, and $c$ are associated with eigenvalues of the field tensor (\ref{F-ar}) (see eq.~(\ref{lbd_eig}) in Section \ref{EvVct}).

As the field strengths which constitute the tensor (\ref{F-ar}) are  static
and spatially uniform, its eigenvectors do not change with points of
Minkowski space. The subspaces spanned by these eigenvectors
constitute foliation of ${\mathbb M}_{\,6}$ by two-dimensional
planes. Traveling from point to point, we construct the coordinate
grid from these planes which covers all the flat spacetime. The
particle's world line may be decomposed into three orbits in these
mutually orthogonal two-dimensional sheets.

We insert matrix (\ref{FL}) into into equations of motion (\ref{Lf-6D}) where kinematic variables in non-linear terms are replaced by the right-hand sides of eqs.~(\ref{adot_FFu}) and (\ref{aa_FFu}). We obtain three pairs of equations defining the orbits in three mutually orthogonal planes: ${\cal M}^{(b)}=\left\{x\in{\mathbb M}_{\,6}|(x^2,x^3,x^4,x^5)=0\right\}$,
${\cal M}^{(a)}=\left\{x\in{\mathbb M}_{\,6}|(x^0,x^1,x^4,x^5)=0\right\}$, and ${\cal M}^{(c)}=\left\{x\in{\mathbb M}_{\,6}|(x^0,x^1,x^2,x^3)=0\right\}$:
\begin{align}
&\left.
\begin{array}{ccc}
\frac{\displaystyle{\rm d}u^0}{\displaystyle{\rm d}\tau}&=&\lambda_b\left(1-\frac{\displaystyle 3\mu}{\displaystyle 2m}\upsilon+\frac{\displaystyle\mu}{\displaystyle m}\lambda_b^2\right)u^1\\[1ex]
\frac{\displaystyle{\rm d}u^1}{\displaystyle{\rm d}\tau}&=&\lambda_b\left(1-\frac{\displaystyle 3\mu}{\displaystyle 2m}\upsilon+\frac{\displaystyle\mu}{\displaystyle m}\lambda_b^2\right)u^0
\end{array}
\right\};\label{1st_pair}\\
&\left.
\begin{array}{ccc}
\frac{\displaystyle{\rm d}u^2}{\displaystyle{\rm d}\tau}&=&\omega_a\left(1-\frac{\displaystyle 3\mu}{\displaystyle 2m}\upsilon-\frac{\displaystyle\mu}{\displaystyle m}\omega_a^2\right)u^3\\[1ex]
\frac{\displaystyle{\rm d}u^3}{\displaystyle{\rm d}\tau}&=&-\omega_a\left(1-\frac{\displaystyle 3\mu}{\displaystyle 2m}\upsilon-\frac{\displaystyle\mu}{\displaystyle m}\omega_a^2\right)u^2
\end{array}
\right\};\label{2nd_pair}\\
&\left.
\begin{array}{ccc}
\frac{\displaystyle{\rm d}u^4}{\displaystyle{\rm d}\tau}&=&\omega_c\left(1-\frac{\displaystyle 3\mu}{\displaystyle 2m}\upsilon-\frac{\displaystyle\mu}{\displaystyle m}\omega_c^2\right)u^5\\[1ex]
\frac{\displaystyle{\rm d}u^5}{\displaystyle{\rm d}\tau}&=&-\omega_c\left(1-\frac{\displaystyle 3\mu}{\displaystyle 2m}\upsilon-\frac{\displaystyle\mu}{\displaystyle m}\omega_c^2\right)u^4
\end{array}
\right\}.\label{3rd_pair}
\end{align}
Here $\lambda_b=(e/m)b$, $\omega_a=(e/m)a$, $\omega_c=(e/m)c$, and
\begin{align}
\upsilon&=\lambda_b^2\left[(u^0)^2-(u^1)^2\right]+\omega_a^2\left[(u^2)^2+(u^3)^2\right]\nonumber\\
&+\omega_c^2\left[(u^4)^2+(u^5)^2\right].
\end{align}
The first pair produces the relation $a^0u^0-a^1u^1=0$. This immediately yields $(u^0)^2-(u^1)^2=const$. The solution to eqs.~(\ref{1st_pair}) defines the hyperbolic orbit in plane $(x^0,x^1)$:
\begin{equation}\label{u0u1}
u^0(\tau)=B\cosh\left(\Lambda_b\tau+\chi_0\right),\,\, u^1(\tau)=B\sinh\left(\Lambda_b\tau+\chi_0\right).
\end{equation}
The frequency
\begin{equation}\label{Lambda_b}
\Lambda_b=\lambda_b\left[1-\frac{3\mu}{2m}\left(\lambda_b^2B^2+\omega_a^2A^2+\omega_c^2C^2\right)+\frac{\mu}{m}\lambda_b^2\right].
\end{equation}
Constant parameter $B$ and phase $\chi_0$ are defined by initial conditions.

The others, eqs.~(\ref{2nd_pair}) and (\ref{3rd_pair}), define circular orbits in $(x^2,x^3)$-plane and in $(x^4,x^5)$-plane:
\begin{align}
u^2(\tau)&=A\sin\left(\Omega_a\tau+\varphi_0\right),\,\, u^3(\tau)=A\cos\left(\Omega_a\tau+\varphi_0\right);\nonumber\\
u^4(\tau)&=C\sin\left(\Omega_c\tau+\phi_0\right),\,\, u^5(\tau)=C\cos\left(\Omega_c\tau+\phi_0\right).
\label{u2-u5}
\end{align}
The frequencies are as follows
\begin{align}
\Omega_a&=\omega_a\left[1-\frac{3\mu}{2m}\left(\lambda_b^2B^2+\omega_a^2A^2+\omega_c^2C^2\right)-
\frac{\mu}{m}\omega_a^2\right]\nonumber\\ \Omega_c&=\omega_c\left[1-\frac{3\mu}{2m}\left(\lambda_b^2B^2+\omega_a^2A^2+\omega_c^2C^2\right)-
\frac{\mu}{m}\omega_c^2\right], \label{Omega_ac}
\end{align}
where constants $A$ and $C$, as well as phases $\varphi_0$ and $\phi_0$, are defined by initial conditions.
Having integrated the solutions with respect to proper time variable we obtain the coordinate functions which define the charge's world line.

Putting either $(a=0, c=0)$ or $(b=0, c=0)$ we obtain the solutions corresponding to field tensors (\ref{FL-b}) or (\ref{FL-a}), respectively.

It is worth noting that the solutions (\ref{u0u1}) and (\ref{u2-u5}) satisfy general equation of motion (\ref{Lf-6D}). The frequencies $\Lambda_b$, $\Omega_a$, and $\Omega_c$ should satisfy the following system of spectral equations:
\begin{equation}\label{system}
\!\!\!\left.
\begin{array}{ccc}
\Lambda_b\left[1+\frac{\displaystyle 3\mu}{\displaystyle 2m}\left(\Lambda_b^2B^2+\Omega_a^2A^2+\Omega_c^2C^2\right)-
\frac{\displaystyle \mu}{\displaystyle m}\Lambda_b^2\right]&=&\lambda_b\\[1em]
\Omega_a\left[1+\frac{\displaystyle 3\mu}{\displaystyle 2m}\left(\Lambda_b^2B^2+\Omega_a^2A^2+\Omega_c^2C^2\right)+
\frac{\displaystyle \mu}{\displaystyle m}\Omega_a^2\right]&=&\omega_a\\[1em]
\Omega_c\left[1+\frac{\displaystyle 3\mu}{\displaystyle 2m}\left(\Lambda_b^2B^2+\Omega_a^2A^2+\Omega_c^2C^2\right)+
\frac{\displaystyle \mu}{\displaystyle m}\Omega_c^2\right]&=&\omega_c
\end{array}
\right\}.
\end{equation}
Crack these equations and you have got the frequencies. (Recall that eqs.~(\ref{Lambda_b}) and (\ref{Omega_ac}) give approximated values.) It is not a trivial matter because the algebraic equations are related to each other. The normalization condition $(u\cdot u)=-1$ yields the relation $B^2-A^2-C^2=1$.

\begin{figure}[ht]
\begin{center}
\includegraphics*[scale=0.8,angle=0,trim=4 4 0 4]{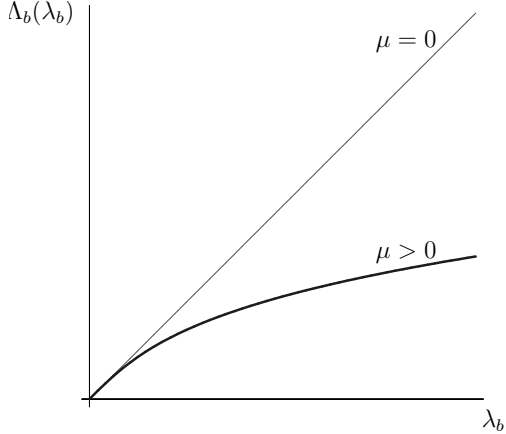}
\end{center}
\caption{Illustration of inertial properties of a rigid charge acted upon a constant electromagnetic field. Solid curve depicts the graph of hyperbolic frequency function (\ref{Ab-psib}) for $\mu>0$.
}\label{Lambda-b}
\end{figure}

Let us consider the motion of a rigid charge in the field of electric type (\ref{FL-b}). The obvious particular solution is $u^\nu(\tau)=(B\cosh(\lambda_b\tau+\chi_0),B\sinh(\lambda_b\tau+\chi_0),
A\sin\varphi_0,A\cos\varphi_0,C\sin\phi_0,C\cos\phi_0)$. The frequency $\Lambda_b$ is the root of the depressed cubic
\begin{equation}\label{eqBl}
\Lambda_b+\frac{\mu}{m}\left(\frac{3}{2}B^2-1\right)\Lambda_b^3=\lambda_b.
\end{equation}
As the parameter $\mu>0$, the only real root can be expressed in terms of hyperbolic functions, i.e. $\Lambda_b=A_b\sinh\psi_b$ where
\begin{equation}\label{Ab-psib}
A_b=\sqrt{\frac{8m}{3\mu\left(3B^2-2\right)}},\, \psi_b=\frac13{\rm arcsinh}\left(\frac{8m\lambda_b}{\mu(3B^2-2)A_b^3}\right).
\end{equation}
If the electromagnetic field is switched off $\lambda_b=0$, then $\Lambda_b=0$. If the parameter $\mu=0$, then $\Lambda_b=\lambda_b$. Therefore, the solution is of true physical sense.

For the field of magnetic type (\ref{FL-a}) the particular solution of the Lorentz-force equation (\ref{Lf-6D}) is $u^\nu(\tau)=(B\cosh(\chi_0),B\sinh(\chi_0),A\sin(\Omega_a\tau+\varphi_0),A\cos(\Omega_a\tau+\varphi_0),C\sin\phi_0,C\cos\phi_0)$. The frequency $\Omega_a$ is the root of the depressed cubic
\begin{equation}\label{eqBom}
\Omega_a+\frac{\mu}{m}\left(\frac{3}{2}A^2+1\right)\Omega_a^3=\omega_a.
\end{equation}
The solution is $\Omega_a=A_a\sinh(\psi_a)$ where amplitude $A_a$ and phase $\psi_a$ are given by eqs.~(\ref{Ab-psib}) where $\omega_a$ and $3A^2+2$ are substituted for constants $\lambda_b$ and $3B^2-2$, respectively.

\section{Algebraic structure of general electromagnetic fields in 6D}\label{Algebr_6D}

An elegant algebraic description of the electromagnetic field is based on its invariants, i.e. on scalar functions of independent components of Faraday tensor (\ref{Far}) which serve as the parameters. The invariants are intimately connected with eigenvalues and eigenvectors of this tensor. For a given square matrix (\ref{F-ar}) an eigenvector $\textbf{u}$ is a non-zero column vector that only change by a scalar $\lambda$ due to linear transformation described by this matrix:
\begin{equation}\label{CharEq}
\hat{F}\textbf{u}=\lambda\textbf{u}.
\end{equation}
Here $\lambda$ is known as the eigenvalue or characteristic root associated with the eigenvector $\textbf{u}$. The scalar is one of roots of characteristic polynomial $p(\lambda)={\rm det}|\hat{F}-\lambda\hat{I}|$:
\begin{equation}\label{eig_v}
\lambda^6+{\cal S}\lambda^4-{\cal Q}\lambda^2-{\cal P}^2=0.
\end{equation}
The coefficients ${\cal S}$, ${\cal Q}$, and ${\cal P}^2$ are the invariants of the electromagnetic field in six dimensions. The coefficients do not depend on the choice of basis in ${\mathbb M}_{\,6}$. Due to the antisymmetry of tensor (\ref{Far}), the vector $\textbf{u}$ is a null-vector. For a stationary and spatially homogeneous field the eigenvectors are the ``bricks'' from which the Lorentz transformation to privileged inertial frame is designed where field strengths are functions of the invariants only.

\subsection{Invariants of electromagnetic field in 6D}\label{Inv}

It is straightforward to see that the invariants are the 0-forms $\!\!\!\phantom{1}^\star({\hat F}\wedge^{\,\star}\!{\hat F})$, $\!\!\!\phantom{1}^\star({\hat F}\wedge{\hat F}\wedge\!\!\!\phantom{1}^\star({\hat F}\wedge{\hat F}))$, and $\!\!\!\phantom{1}^\star({\hat F}\wedge{\hat F}\wedge{\hat F})$. Symbols $\wedge$ and $\!\!\!\phantom{1}^\star$ denote the wedge product and the Hodge star operator, respectively. We are manipulating differential forms with the help of rules presented in the Handbook \cite{Schutz}.

We define 2-form $\hat{G}=\!\!\!\phantom{1}^{\,\star}({\hat F}\wedge{\hat F})$ as the Hodge dual of the wedge product of the Faraday 2-form ${\hat F}=\frac12F_{\alpha\beta}\,{\rm d}x^\alpha\wedge{\rm d}x^\beta$ on itself. In terms of the Minkowski rectangular coordinates the form  ${\hat G}=\frac12G_{\alpha\beta}\,{\rm d}x^\alpha\wedge{\rm d}x^\beta$ is specified by the skew symmetric matrix
\begin{equation}\label{Gar}
(G_{\alpha\beta})=\left(
\begin{array}{cccccc}
0&B^1&B^2&B^3&B^4&B^5\\
-B^1&0&G_{12}&-G_{31}&G_{14}&-G_{51}\\
-B^2&-G_{12}&0&G_{23}&-G_{42}&G_{25}\\
-B^3&G_{31}&-G_{23}&0&G_{34}&-G_{53}\\
-B^4&-G_{14}&G_{42}&-G_{34}&0&G_{45}\\
-B^5&G_{51}&-G_{25}&G_{53}&-G_{45}&0
\end{array}
\right).
\end{equation}
The elements of this matrix involve the magnetic field five-vector which is specified by components $(B^1,B^2,B^3,B^4,B^5)$ where $B^i=G_{0i}$. For example, $B^1=F_{23}F_{45}-F_{42}F_{53}+F_{25}F_{34}$. More generally,
\begin{equation}\label{G-ab}
G_{\alpha\beta}=\frac{1}{(2!)^3}\varepsilon_{\alpha\beta\gamma\delta\mu\nu}F^{\gamma\delta}F^{\mu\nu},
\end{equation}
where the components $F^{\mu\nu}=\eta^{\mu\alpha}\eta^{\nu\beta}F_{\alpha\beta}$ define 2-vector $\textbf{F}=\frac12F^{\alpha\beta}\,\textbf{e}_\alpha\wedge\textbf{e}_\beta$ where $(\textbf{e}_\alpha, \alpha=\overline{0,5})$ is the basis of for Minkowski space ${\mathbb M}_{\,6}$ and $({\rm d}x^\alpha, \alpha=\overline{0,5})$ is the dual basis of 1-forms. The Levi-Civita symbol $\epsilon_{\alpha\beta\gamma\delta\mu\nu}$ is a tensor of rank six and is defined by $0$ if any two labels are the same, $+1$ if $\alpha,\beta,\gamma,\delta,\mu,\nu$ is an even permutation of $0,1,2,3,4,5$, and $-1$ if the set of labels is an odd permutation of these numbers. The Levi-Civita symbol is anti-symmetric on each pair of indexes.
The space-space elements of matrix $\hat{G}$ contain components of electric field, for example $G_{12}=E^3F_{45}+E^4F_{53}+E^5F_{34}$.

Direct calculations of ${\rm det}|\hat{F}-\lambda\hat{I}|$ result the following coefficients of characteristic polynomial (\ref{eig_v}):
\begin{eqnarray}
{\cal S}&=&-4\!\!\phantom{1}^\star({\hat F}\wedge^{\,\star}\!{\hat F})\nonumber\\
&=&\frac{1}{4!}\varepsilon_{\alpha\beta\gamma\delta\mu\nu}\frac12F_{\alpha\beta}\,
\varepsilon_{\kappa\sigma\gamma\delta\mu\nu}\frac12F^{\kappa\sigma}\nonumber\\
&=&-\sum_{i=1}^5(E^i)^2+\sum_{i<k}F_{ik}^2,\label{S-inv}
\end{eqnarray}
\begin{eqnarray}
{\cal Q}&=&4\!\!\phantom{1}^\star({\hat F}\wedge{\hat F}\wedge{\hat G})\nonumber\\
&=&\frac12\varepsilon_{\alpha\beta\gamma\delta\mu\nu}\frac12F^{\alpha\beta}\frac12F^{\gamma\delta}\frac12G^{\mu\nu}\,
\nonumber\\
&=&-\sum_{i=1}^5(B^i)^2+\sum_{i<k}G_{ik}^2,\label{Q-inv}
\end{eqnarray}
\begin{eqnarray}
{\cal P}&=&\frac43\!\!\phantom{1}^\star({\hat F}\wedge{\hat F}\wedge{\hat F})\nonumber\\
&=&\frac16\varepsilon_{\alpha\beta\gamma\delta\mu\nu}\frac12F^{\alpha\beta}\frac12F^{\gamma\delta}\frac12F^{\mu\nu}\nonumber\\
&=&\sum_{i=1}^5E^iB^i.\label{P-inv}
\end{eqnarray}
Restriction of the source 6-forms to Minkowski space of four dimensions ${\mathbb M}_{\,4}\subset{\mathbb M}_{\,6}$ and application of the Hodge star operator gives the well-known ``four-dimensional'' invariants
\begin{equation}\label{SQ-M4}
\left.{\cal S}\right|_{{\mathbb M}_4}=-\mathbf{E}^2+\mathbf{B}^2,\quad \left.{\cal Q}\right|_{{\mathbb M}_4}=(\mathbf{E}\mathbf{B})^2,
\end{equation}
where $\mathbf{E}$ is the electric field 3-vector and $\mathbf{B}$ is the magnetic field 3-vector. Restriction of invariant (\ref{P-inv}) to four-dimensional flat spacetime results zero: $\left.{\cal P}\right|_{{\mathbb M}_4}=0$.

Inspired by eqs.~(\ref{SQ-M4}), we express the ``six-dimensional'' invariants in the concise form:
\begin{equation}\label{SQP-M6}
{\cal S}=-\mathbf{E}^2+{\cal B},\quad {\cal Q}=-\mathbf{B}^2+{\cal E},\quad {\cal P}=(\mathbf{E}\mathbf{B}),
\end{equation}
where the calligraphic letter ${\cal B}$ denotes the sum of ten squared components $F_{ik}^2$ and ${\cal E}$ is the sum of ten squared terms $G_{ik}^2$.

\subsection{Domain of invariants}\label{DomainInv}

We substitute $w$ for $\lambda^2$ in the characteristic polynomial (\ref{eig_v}):
\begin{equation}\label{pol-cubic}
w^3+{\cal S}w^2-{\cal Q}w-{\cal P}^2=0.
\end{equation}
Vieta's formulae relate the coefficients of this cubic polynomial and its roots $w_0$, $w_1$, and $w_2$:
\begin{eqnarray}
w_0+w_1+w_2&=&-{\cal S},\label{Vieta_S}\\
w_0w_1+w_0w_2+w_1w_2&=&-{\cal Q},\label{Vieta_q}\\
w_0w_1w_2&=&{\cal P}^2.\label{Vieta_p2}
\end{eqnarray}
The coefficients are the functions (\ref{S-inv})-(\ref{P-inv}) of the components of the Faraday tensor (\ref{Far}) which serve as the parameters. In this Paragraph we establish their domain.

Cardano's discriminant
\begin{equation}\label{Cardano}
\Delta=18{\cal S}{\cal Q}{\cal P}^2+4{\cal S}^3{\cal P}^2+{\cal S}^2{\cal Q}^2+4{\cal Q}^3-27{\cal P}^4
\end{equation}
determine the type of roots of the algebraic equation (\ref{pol-cubic}). The values of invariants of electromagnetic field ${\cal S}$, ${\cal Q}$, and ${\cal P}$ do not depend on the choice of inertial frame of reference where the electromagnetic field strengths are measured. When considering the general case ${\cal P}\neq 0$, it is convenient to use the Lorentz frame where the electric field five-vector ${\bf E}$ and the magnetic field five-vector ${\bf B}$ are collinear. In this specific reference frame the invariants (\ref{SQP-M6}) take the form
\begin{equation}\label{InvSpecRF}
{\cal S}=-(E^1)^2+{\cal B}',\;\, {\cal Q}=-(B^1)^2+(E^1)^2{\cal B}',\;\, {\cal P}=E^1B^1,
\end{equation}
where ${\cal B}'=(F_{23})^2+(F_{42})^2+(F_{25})^2+(F_{34})^2+(F_{53})^2+(F_{45})^2$. Inserting these into eq.~(\ref{Cardano}), canceling like terms, and factoring we express the Cardano's discriminant in the form
\begin{eqnarray}\label{Cardan}
\Delta&=&\left[\left(F_{23}-F_{45}\right)^2+\left(F_{25}-F_{34}\right)^2+\left(F_{42}+F_{53}\right)^2\right]\nonumber\\
&\times&\left[\left(F_{23}+F_{45}\right)^2+\left(F_{25}+F_{34}\right)^2+\left(F_{42}-F_{53}\right)^2\right]\nonumber\\
&\times&\left[\left(E^1\right)^2{\cal B}'+\left(E^1\right)^4+\left(B^1\right)^2\right].
\end{eqnarray}
Since $\Delta>0$, the equation (\ref{pol-cubic}) has three real distinct roots.

The invariants ${\cal S}$ and ${\cal Q}$ can not be negative simultaneously. If ${\cal S}<0$ then $(E^1)^2>{\cal B}'$ and, therefore, ${\cal Q}=-(B^1)^2+(E^1)^2{\cal B}'>-(B^1)^2+({\cal B}')^2$. Factoring this expression we obtain
$$
-(B^1)^2+({\cal B}')^2=\left(f_{23}^-+f_{53}^++f_{34}^-\right)\left(f_{23}^++f_{53}^-+f_{34}^+\right),
$$
where all the terms $f_{23}^{\pm}=F_{23}^2 \pm F_{23}F_{45}+F_{45}^2$, $f_{53}^{\pm}=F_{53}^2 \pm F_{53}F_{42}+F_{42}^2$, and $f_{34}^{\pm}=F_{34}^2 \pm F_{34}F_{25}+F_{25}^2$ are non-negative. Consequently, the condition ${\cal S}<0$ yields ${\cal Q}>0$. If we choose ${\cal S}>0$, then ${\cal Q}$ can be either negative or positive. Taking into account the Vieta's formulae
(\ref{Vieta_S})-(\ref{Vieta_p2}) we conclude that the characteristic polynomial (\ref{pol-cubic}) possesses one positive root, say $w_0$, and two negative ones, $w_1$ and $w_2$.

\begin{figure}[ht]
\begin{center}
\includegraphics*[scale=0.89,angle=0,trim=4 4 0 4]{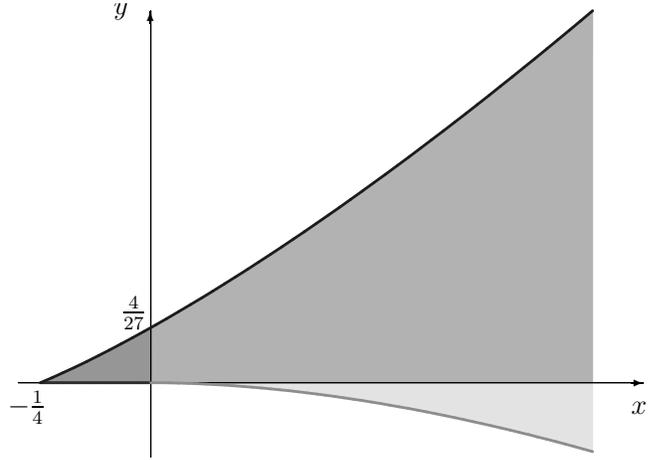}
\end{center}
\caption{Domain of Cardano's discriminant (\ref{Cardxy}). The shaded area pictures the region of possible values of field's invariants. It is bounded by the curve $y_+(x)$, $x\in[-1/4,+\infty[$, the interval $[-1/4,0]$ on the $x$-axis, and the curve $y_-(x)$ where $x\in[0,+\infty[$.  
}\label{Cardano-Dm}
\end{figure}

Our next task is to establish the domain of invariants of electromagnetic field in six dimensions. To visualize the tree-dimensional domain of function (\ref{Cardano}) we reduce it to a flat map. We pass to the dimensionless variables
\begin{equation}\label{Dmnsless}
\omega=\frac{w}{|{\cal S}|},\qquad x=\frac{{\cal Q}}{{\cal S}^2},\qquad y=\frac{{\cal P}^2}{{\cal S}^3}.
\end{equation}
The polynomial (\ref{pol-cubic}) simplifies
\begin{equation}\label{xy-cubic}
\omega^3+{\rm sqn}({\cal S})\omega^2-x\omega-{\rm sqn}({\cal S})y=0.
\end{equation}
The transformed Cardano's discriminant $\Delta'=\Delta/{\cal S}^6$ becomes
\begin{eqnarray}\label{Cardxy}
\Delta'&=&18xy+4y+x^2+4x^3-27y^2\nonumber\\
&=&27\left[\phantom{1^1}\!\!\!\!\!y_+(x)-y\right]\left[y-y_-(x)\phantom{1^1}\!\!\!\!\!\right],
\end{eqnarray}
where
\begin{equation}\label{y_pm}
y_\pm(x)=\frac{2}{27}\left[1+\frac92x\pm\left(1+3x\right)^{3/2}\right].
\end{equation}
The domain is the area in the $(xy)$-plane bounded by the curves $y_-(x)$ and $y_+(x)$ (see Fig.~\ref{Cardano-Dm}).

The limiting curves $y_+(x)$ and $y_-(x)$ are projections of the two-dimensional surface
\begin{equation}\label{P-2}
{\cal P}^2=\frac{2}{27}\left[{\cal S}^3+\frac92{\cal Q}{\cal S}+\left({\cal S}^2+3{\cal Q}\right)^{3/2}\right].
\end{equation}
on the regions $S^+=\left\{({\cal Q},{\cal S})\in\mathbb{R}^2|{\cal Q}>-{\cal S}^2/4,{\cal S}>0\right\}$ and
$S^-=\left\{({\cal Q},{\cal S})\in\mathbb{R}^2|{\cal Q}>0,{\cal S}<0\right\}$, respectively. On the limiting surface (\ref{P-2}) the Cardano's discriminant (\ref{Cardano}) vanishes. The cubic polynomial (\ref{pol-cubic}) has a simple root and a double root in this case. The regions $S^+$ and $S^-$ are depicted in Fig.~\ref{QS-plane}.

\subsection{Classification of fields}\label{ClassF}

We base our classification of electromagnetic fields in six dimensions on the analysis of the characteristic polynomial (\ref{eig_v}). Fields are divided into classes according to whether ${\cal P}\ne 0$ or ${\cal P}=0$ as well as whether ${\cal S}<0$ or ${\cal S}>0$. The other invariant is ${\cal Q}>-{\cal S}^2/4$.

\begin{itemize}
\item[(A)] ${\cal P}\neq 0.$ There are three distinct real roots of the reduced characteristic polynomial, one positive and two negative. The set of eigenvalues consists of three pairs, one pair of real numbers, and two pairs of pure imaginary numbers.
\item[(B)] ${\cal P}=0,\,{\cal Q}\ne 0.$ There is one zero root of the reduced characteristic polynomial.

Since the sign of ${\cal Q}$ is invariant, there are two possibilities:
\begin{itemize}
\item[(a)]
If ${\cal Q}>0$, there are two non-trivial roots of the reduced characteristic polynomial, positive and negative. The set of eigenvalues consists of double degenerate non-defective zero and two pairs, one real and one pure imaginary. Such a field has four-dimensional analogue which belongs to the set of {\it crossed} fields.
\item[(b)]
If $-{\cal S}^2/4<{\cal Q}<0$, there are two negative non-trivial roots of the reduced characteristic polynomial. The set of eigenvalues consists of double degenerate non-defective zero and two pairs of pure imaginary numbers. The field have not analogue in conventional electrodynamics.
\end{itemize}
\item[(C)] ${\cal P}=0,\, {\cal Q}=0,\, {\cal S}\neq 0.$ There are double degenerate zero and one non-trivial real root of the reduced characteristic polynomial. Characteristic polynomial has fourth degenerate zero eigenvalue which can be either defective or non-defective.

    If zero is non-defective eigenvalue, there are two possibilities:
\begin{itemize}
\item[(a)]
If ${\cal S}<0$, the non-trivial root of the reduced characteristic polynomial is positive. The set of eigenvalues consists of fourth degenerate zero and pair of real numbers, negative and positive. Such a field has four-dimensional analogue which is said to be of {\it electric type}.
\item[(b)]
If ${\cal S}>0$, the non-trivial root of the reduced characteristic polynomial is negative. The set of eigenvalues consists of fourth degenerate zero and pair of pure imaginary numbers. Such a field has four-dimensional analogue which is called {\it magnetic type} field.
\end{itemize}

If zero is defective eigenvalue, the field matrix in non-diagonalizable. The Jordan normal form consists of two Jordan blocks: one-dimensional zero block and $3\times 3$ matrix with zeroes on the main diagonal and ones on the superdiagonal.

\item[(D)] ${\cal P}=0,\, {\cal Q}=0,\, {\cal S}=0.$

If all the ``six-dimensional'' invariants are equal to zero, we deal with the {\it null} field.
\end{itemize}

\subsection{Eigenvalues}\label{EvVct}

Let us consider general case (A) ${\cal P}\ne 0$. The subset of domain od discriminant (\ref{Cardano}) consists of all the points on the map projection Fig.~\ref{Cardano-Dm}, excepting the ray $x\in [-1/4,+\infty[$ on the abscissa axis. The roots of  cubic polynomial (\ref{pol-cubic}) has three distinct real roots which can be expressed in terms of trigonometric functions
\begin{equation}\label{b2}
w_k=-\frac13{\cal S}+\frac23\sqrt{{\cal S}^2+3{\cal Q}}\cos\left(\psi+\frac{2\pi}{3}k\right).
\end{equation}
The angle $\psi$ is given by
\begin{equation}\label{psi}
\psi({\cal S},{\cal Q},{\cal P})=\frac13\arccos\left(\frac{\frac{27}{2}{\cal P}^2-{\cal S}^3-\frac92{\cal Q}{\cal S}}{\left({\cal S}^2+3{\cal Q}\right)^{3/2}}\right).
\end{equation}
The roots are ordered as $w_1<w_2<w_0$. The largest root $w_0=b^2$ is positive, the others are negative: $w_1=-a^2$ and $w_2=-c^2$.

The set of eigenvalues consists of three pairs, one real and the others pure imaginary
\begin{equation}\label{lbd_eig}
\{\lambda\}=\{+b,-b,+{\rm i}a,-{\rm i}a,+{\rm i}c,-{\rm i}c\}.
\end{equation}
Any characteristic root from this set is associated with corresponding eigenvector being the solution of matrix equation (\ref{CharEq}). In Appendix \ref{PrivRF} we construct a square matrix $U$ whose columns are the six linearly independent eigenvectors ranged according to the range of eigenvalues in the list (\ref{lbd_eig}). As the matrix invertible, we diagonalize the field tensor $F^\alpha{}_\beta=\eta^{\alpha\mu}F_{\mu\beta}$ as follows
$$
L=U^{-1}\hat{F}U.
$$
The matrix $L$ is composed from the eigenvalues (\ref{lbd_eig}) on the diagonal. However, the diagonal matrix does not describe an electromagnetic field. To build the electromagnetic field tensor we modify the transformation to eigenbasis by additional linear transformation defined by the matrix
\begin{equation}\label{J}
(J_{\alpha\beta})=\frac{1}{\sqrt{2}}\left(
\begin{array}{cccccc}
1&1&0&0&0&0\\
-1&1&0&0&0&0\\
0&0&{\rm i}&1&0&0\\
0&0&-{\rm i}&1&0&0\\
0&0&0&0&{\rm i}&1\\
0&0&0&0&-{\rm i}&1
\end{array}
\right).
\end{equation}
We obtain the electromagnetic field tensor $\hat{F}'=J^{-1}LJ$ which depends on the field's invariants only (see eq. (\ref{FL})).

Looking at the chain of transformations
\begin{eqnarray}
\hat{F}'&=&J^{-1}LJ\nonumber\\
&=&J^{-1}U^{-1}\hat{F}UJ\nonumber\\
&=&\Lambda^{-1}\hat{F}\Lambda\label{Fprime}
\end{eqnarray}
we see that the tensor (\ref{FL}) defines the electromagnetic field in the privileged reference frame which is related to the initial inertial frame by the Lorentz matrix $\Lambda=UJ$ which is presented in Appendix \ref{PrivRF}.


\subsubsection{${\cal P}=0, {\cal Q}\neq 0$: $({\cal Q}, {\cal S})$ plane}
\begin{figure}[ht]
\begin{center}
\includegraphics*[scale=0.8,angle=0,trim=4 4 0 4]{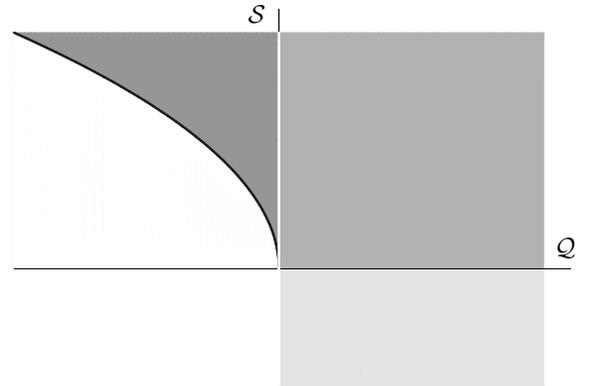}
\end{center}
\caption{Domain of Cardano's discriminant (\ref{Cardano}) in $({\cal Q}{\cal S})$-plane. The shaded area pictures the region of possible values of field's invariants. It is bounded by the curve ${\cal Q}=-{\cal S}^2/4$ where ${\cal S}\in[0,+\infty[$, and the negative ordinate half-axis.  
}\label{QS-plane}
\end{figure}

If the electric field and the magnetic field are mutually orthogonal in a given inertial frame, they are orthogonal in any other frame of reference. If ${\cal P}=0$. In the projection map Fig.~\ref{Cardano-Dm} the points lie on the ray $x\geq -1/4$, $y=0$. It is worth noting that the coordinate origin is the punctured point. The ray corresponds to the shaded region in the $({\cal Q}, {\cal S})$ plane which is depicted in Fig.~\ref{QS-plane}.

If ${\cal P}=0$, the cubic polynomial (\ref{pol-cubic}) simplifies:
\begin{equation}\label{pol-squared}
w\left(w^2+{\cal S}w-{\cal Q}\right)=0.
\end{equation}
The roots are $0$ and $w_\pm=\left(-{\cal S}+\sqrt{{\cal S}^2+4{\cal Q}}\right)/2$. So far as arranging is concerning, there are two possibilities.

\paragraph{${\cal Q}>0$.}\label{Bot-edge-1qu} The roots are arranged as follows:
\begin{align}
\left.w_0({\cal S},{\cal Q},{\cal P})\right|_{{\cal P}=0}&=w_+({\cal S},{\cal Q}),\nonumber\\
\left.w_2({\cal S},{\cal Q},{\cal P})\right|_{{\cal P}=0}&=0,\nonumber\\
\left.w_1({\cal S},{\cal Q},{\cal P})\right|_{{\cal P}=0}&=w_-({\cal S},{\cal Q}).\nonumber
\end{align}
The spectrum consists of two pairs, one real and the other pure imaginary, and double degenerate zero
\begin{equation}\label{lbd_eigP0Qpl}
\{\lambda\}=\{+b,-b,+{\rm i} a,-{\rm i} a,0,0\},
\end{equation}
where $b=\sqrt{w_+({\cal S},{\cal Q})}$ and $a=\sqrt{-w_-({\cal S},{\cal Q})}$.
This electromagnetic field has analogue in conventional electrodynamics. In the limiting case (\ref{SQ-M4}) we obtain the so-called ``crossed field''. In the privileged inertial frame the electromagnetic field tensor simplifies
\begin{equation}\label{FLP0Qpl}
(F^\alpha{}_\beta)'=\left(
\begin{array}{cccccc}
0&b&0&0&0&0\\
b&0&0&0&0&0\\
0&0&0&a&0&0\\
0&0&-a&0&0&0\\
0&0&0&0&0&0\\
0&0&0&0&0&0
\end{array}
\right).
\end{equation}

\paragraph{$-{\cal S}^2/4<{\cal Q}<0$, ${\cal S}>0$.}\label{Bot-edge-2qu}
Both the non-trivial roots are negative:
\begin{align}
\left.w_0({\cal S},{\cal Q},{\cal P})\right|_{{\cal P}=0}&=0,\nonumber\\
\left.w_2({\cal S},{\cal Q},{\cal P})\right|_{{\cal P}=0}&=w_+({\cal S},{\cal Q}),\nonumber\\
\left.w_1({\cal S},{\cal Q},{\cal P})\right|_{{\cal P}=0}&=w_-({\cal S},{\cal Q}).\nonumber
\end{align}
The spectrum consists of double degenerate zero and two pairs, both pure imaginary
\begin{equation}\label{lbd-eigP0Qmn}
\{\lambda\}=\{0,0,+{\rm i} a,-{\rm i} a,+{\rm i} c,-{\rm i} c\},
\end{equation}
where $a=\sqrt{-w_+({\cal S},{\cal Q})}$ and $c=\sqrt{-w_-({\cal S},{\cal Q})}$. This electromagnetic field has no analogue in conventional electrodynamics. In the privileged inertial frame the electromagnetic field tensor takes the form
\begin{equation}\label{FLP0Qmn}
(F^\alpha{}_\beta)''=\left(
\begin{array}{cccccc}
0&0&0&0&0&0\\
0&0&0&0&0&0\\
0&0&0&a&0&0\\
0&0&-a&0&0&0\\
0&0&0&0&0&c\\
0&0&0&0&-c&0
\end{array}
\right).
\end{equation}
The Lorentz matrices transforming to fields (\ref{FLP0Qpl}) and (\ref{FLP0Qmn}) are presented in Appendix \ref{PrivRF}.

\subsubsection{${\cal P}=0, {\cal Q}= 0$: ${\cal S}$-line}

If both the coefficients ${\cal Q}$ and ${\cal P}$ in the characteristic polynomial (\ref{eig_v}) vanish, the spectrum contains fourth degenerate zero:
\begin{equation}\label{eig_v00}
\lambda^4\left(\lambda^2+{\cal S}\right)=0.
\end{equation}
It contains also the pair of distinct eigenvalues, either real $\pm\sqrt{-{\cal S}}$ if ${\cal S}<0$ or pure imaginary $\pm{\rm i}\sqrt{{\cal S}}$ if ${\cal S}>0$.

Direct calculations show that the rank of the electromagnetic field matrix (\ref{F-ar}) is
$$
\left.rank(F)\right|_{{\cal Q}=0,{\cal P}=0}=4.
$$
The geometric multiplicity is less than the algebraic one. It is easy to show that
$$
\left.rank(F^2)\right|_{{\cal Q}=0,{\cal P}=0}=3,\quad \left.rank(F^3)\right|_{{\cal Q}=0,{\cal P}=0}=2,
$$
and ranks of the matrix $F$ in higher powers no longer decrease. Therefore, zero is defective eigenvalue.  The Jordan normal form consists of two Jordan blocks: one-dimensional zero block and $3\times 3$ matrix with zeroes on the main diagonal and ones on the superdiagonal. Depending on the sign of the only non-trivial invariant ${\cal S}$, the field matrix in the eigenbasis takes the form, either
\begin{equation}\label{J011-min}
\left.\phantom{1^1}L\right|_{{\cal S}<0}=\left(
\begin{array}{cccccc}
+b&0&0&0&0&0\\
0&-b&0&0&0&0\\
0&0&0&1&0&0\\
0&0&0&0&1&0\\
0&0&0&0&0&0\\
0&0&0&0&0&0
\end{array}
\right),
\end{equation}
or
\begin{equation}\label{J011-plus}
\left.\phantom{1^1}L\right|_{{\cal S}>0}=\left(
\begin{array}{cccccc}
0&1&0&0&0&0\\
0&0&1&0&0&0\\
0&0&0&0&0&0\\
0&0&0&0&0&0\\
0&0&0&0&+{\rm i}c&0\\
0&0&0&0&0&-{\rm i}c
\end{array}
\right).
\end{equation}
Here $b=\sqrt{-{\cal S}}$ and $c=\sqrt{{\cal S}}$. Such an electromagnetic field has no limit in four dimensions. We are interested in the diagonalizable matrices because diagonal matrices have equivalents in four dimensions. In this case zero eigenvalue has four-dimensional eigenspace and the Jordan normal form is $4\times 4$ matrix in which all the entries are zeroes. Besides ${\cal P}=0$ and ${\cal Q}=0$, additional conditions on Faraday tensor (\ref{F-ar}) are necessary which provide $rank(F)=2$.

To establish them we pay attention to the matrix $G^\mu{}_\beta=\eta^{\mu\alpha}G_{\alpha\beta}$ where $G_{\alpha\beta}$ is given by eq.~(\ref{Gar}). We subtract $\Lambda$ from the diagonal to find $G-\Lambda\mathbb{I}$ and calculate its determinant. The characteristic polynomial must be zero:
\begin{equation}\label{eig_L}
\Lambda^6+{\cal Q}\Lambda^4-{\cal S}{\cal P}^2\Lambda^2-{\cal P}^4=0.
\end{equation}
Putting $\Lambda^2=W$ we transform it in the cubic polynomial. Its Cardano's discriminant is proportional to the discriminant (\ref{Cardano}) of the characteristic polynomial (\ref{eig_v}): $D={\cal P}^4\Delta$. If the invariant ${\cal P}\neq 0$, there are three real distinct roots, one positive and the others negative. Trivial analysis of corresponding Vieta's relations yields $$
W_0=w_1w_2,\quad W_1=w_0w_2,\quad W_2=w_0w_1,
$$
where the small letters denote the roots of cubic polynomial (\ref{pol-cubic}).

If the invariant ${\cal P}=0$, the matrix $F$ becomes degenerate: $rank(F)=4$. The conjugated matrix (\ref{Gar}) is degenerate too: $rank(G)=rank(F)-2$. To illustrate the situation we consider the field tensor of rank $4$:
\begin{equation}\label{Far-M4}
(F^\mu{}_\nu)=\left(
\begin{array}{cccccc}
0&E^1&E^2&E^3&0&0\\
E^1&0&F_{12}&-F_{31}&0&0\\
E^2&-F_{12}&0&F_{23}&0&0\\
E^3&F_{31}&-F_{23}&0&0&0\\
0&0&0&0&0&0\\
0&0&0&0&0&0
\end{array}
\right).
\end{equation}
Its Hodge dual counterpart
\begin{equation}\label{Gar-M4}
(G^\mu{}_\nu)=\left(
\begin{array}{cccccc}
0&0&0&0&0&0\\
0&0&0&0&0&0\\
0&0&0&0&0&0\\
0&0&0&0&0&0\\
0&0&0&0&0&G_{45}\\
0&0&0&0&-G_{45}&0
\end{array}
\right).
\end{equation}
is of rank $2$.

If both the invariants ${\cal Q}$ and ${\cal P}$ vanish, the spectrum equation (\ref{eig_v}) admits fourth degenerate zero while the spectrum equation (\ref{eig_L}) becomes simple $\Lambda^6=0$. Imposing the condition $G_{45}\equiv E^1F_{23}+E^2F_{31}+E^3F_{12}=0$ on the electromagnetic field (\ref{Far-M4}) we obtain the matrix which has rank $2$. All the elements of the conjugated matrix (\ref{Gar-M4}) are identically equal to zero in this case.

To construct the field matrix (\ref{F-ar}) of rank $2$ we should provide resetting to zero of elements of the matrix (\ref{Gar}). It is sufficient to equate to zero the elements $G_{ij}, G_{ki}, G_{jk}$ and $B^i, B^j, B^k$, and solve the algebraic equations with respect to the elements $F_{ij}, F_{ki}, F_{jk}$ and $E^i, E^j, E^k$ of matrix (\ref{F-ar}). There are ten triples $(ijk)$ if the indices run from $1$ to $5$, e.g.
\begin{eqnarray}
E^1&=&-\frac{E^4F_{51}+E^5F_{14}}{F_{45}},\quad F_{12}=\frac{F_{14}F_{25}-F_{51}F_{42}}{F_{45}}\nonumber\\
E^2&=&\frac{E^4F_{25}+E^5F_{42}}{F_{45}},\quad F_{31}=\frac{F_{14}F_{53}-F_{51}F_{34}}{F_{45}},\nonumber\\
E^3&=&-\frac{E^4F_{53}+E^5F_{34}}{F_{45}},\quad F_{23}=\frac{F_{42}F_{53}-F_{25}F_{34}}{F_{45}}.\nonumber
\end{eqnarray}
Diagonalization of this matrix yields the Jordan normal form being $4\times 4$ matrix in which all the entries are zeroes.

If ${\cal S}<0$, the primed electromagnetic field tensor has the form (\ref{FL-b}). Such a field is said to be of {\it electric type}. If ${\cal S}>0$, the field tensor is given by eq. (\ref{FL-a}). Its analogue in the conventional electrodynamics is called {\it magnetic type} field. With the help of additional linear transformations they are rearranged into skew symmetric matrices describing electromagnetic fields in specific inertial frames (see Appendix \ref{PrivRF}).

\section{Compactification of two extra dimensions}\label{Compact6-4}

Our task in this Section is to reduce the dimensionality of six-dimensional Minkowski space and pass to the ordinary electrodynamics in four dimensions. We shall perform the procedure in two steps. At first we map the flat spacetime $\mathbb{M}_6$ onto the five-dimensional hyperboloid $\mathbb{H}_5$
\begin{equation}\label{deS-H5}
\eta_{AB}\chi^A\chi^B=\frac{1}{H^2},
\end{equation}
and then repeat the surjection via passing to the four-dimensional hyperboloid in $\mathbb{H}_5$. The hyperboloid represents the de Sitter spacetime which is the solution of the Einstein field equation with cosmological constant $\Lambda$. The constant $\Lambda=3H^2$ is related to the inverse length parameter involved in eq.~(\ref{deS-H5}). The de Sitter space with constant curvature ${\cal R}=4\Lambda$ has ten Killing vectors, i.e. is maximally symmetric. The parameter $H^2$ is involved in the de Sitter group governing the kinematics in the de Sitter space. In the cosmological parameter limit $\Lambda\to 0$, the de Sitter group contracts to the Poincar\'{e} group governing the kinematics in the Minkowski space $\mathbb{M}_4$. Indeed, $\Lambda=0$ means the absence of gravitation: the flat spacetime is a solution of the sourceless Einstein equation.

To parameterize the points of hyperboloid (\ref{deS-H5}) we use the five-dimensional generalization of stereographic coordinates \cite{Aldro07}
\begin{align}
\chi^a&=\Omega(X)X^a,\nonumber\\
\chi^5&=-\frac{1}{H}\Omega(X)\left(1-\frac{H^2}{4}\Sigma^2\right),\label{Trans6-5}
\end{align}
where
\begin{equation}
\Omega(X)=\left(1+\frac{H^2}{4}\Sigma^2\right)^{-1}, 
\quad\Sigma^2=\eta_{ab}X^aX^b. \label{Sigma4-2}
\end{equation}
Small Latin indices run from $0$ to $4$. Further we map the 5-dimensional hyperboloid $\mathbb{H}_5$ onto the 4-dimensional hyperboloid $\mathbb{H}_4$ defined by the equation
\begin{equation}\label{deS-H4}
\eta_{ab}X^aX^b=\frac{1}{h^2}.
\end{equation}
It is parameterized by the four-dimensional stereographic coordinates \cite{Aldro07}:
\begin{align}
X^\alpha&=\omega(x)x^\alpha,\nonumber\\
X^4&=-\frac{1}{h}\omega(x)\left(1-\frac{h^2}{4}\sigma^2\right),\label{Trans5-4}
\end{align}
where
\begin{equation}
\omega(x)=\left(1+\frac{h^2}{4}\sigma^2\right)^{-1}, 
\quad\sigma^2=\eta_{\alpha\beta}x^\alpha x^\beta. \label{sigma3-2}
\end{equation}
Small Greek indices run from $0$ to $3$. Composition of these surjections is written as
\begin{align}
\chi^\alpha&=\Omega(\kappa)\omega(x)x^\alpha,\nonumber\\
\chi^4&=-\frac{1}{h}\Omega(\kappa)\omega(x)\left(1-\frac{h^2}{4}\sigma^2\right),\nonumber\\
\chi^5&=-\frac{1}{H}\frac{1-\kappa^2}{1+\kappa^2},\label{Trans6-4}
\end{align}
where $\Omega(\kappa)=\left(1+\kappa^2\right)^{-1}$ depends on the dimensionless constant $\kappa=H/(2h)$. The de Sitter metric on $\mathbb{H}_4$ is induced by the metric in the flat spacetime $\mathbb{M}_6$:
\begin{align}
{\rm d}s^2&=\left.\eta_{AB}{\rm d}\chi^A{\rm d}\chi^B\right|_{\mathbb{H}_4}\nonumber\\
&=\Omega^2(\kappa)\omega^2(x)\eta_{\alpha\beta}{\rm d}x^\alpha{\rm d}x^\beta. \label{metric6-4}
\end{align}
Let us define the {\it cosmological constant limit} when both the constants $H$ and $h$ converge to 0 simultaneously, while their fraction $\kappa$ remains a finite constant. In this assumption the function $\omega(x)$ in eq.~(\ref{metric6-4}) is equal to $1$. To obtain the standard flat space metric we scale the Minkowski coordinates $\Omega(\kappa)x^\alpha\to x^\alpha$.

Now restricting the electromagnetic field two-form (\ref{Far}) to the de Sitter space ${\mathbb{H}_4}$ we obtain the field 2-form in four dimensions
\begin{align}
\hat{f}&=\frac12\Omega^2(\kappa)\omega^2(x)\left[\phantom{\frac11}\!\!\!\!F_{\alpha\beta}+h\omega(x)\left(x_\alpha F_{4\beta}+F_{\alpha 4} x_\beta\right)\right.\nonumber\\
&\left.-\frac{h^2}{2}\omega(x)\left(x_\alpha F_{\nu\beta}x^\nu+F_{\alpha\nu}x^\nu x_\beta\right)\right]{\rm d}x^\alpha\wedge{\rm d}x^\beta.
\end{align}
In the cosmological constant limit it takes the form $\frac12F_{\alpha\beta}{\rm d}x^\alpha\wedge{\rm d}x^\beta$ where we use the scaled Minkowski coordinates.

In six dimensions the particle dynamics is governed by the action (\ref{Spart}). In order to restrict the dynamics on the hyperboloid one should take into account two holonomic constraints: $z^5=const$ and $\eta_{ab}z^az^b=const$. Recall that small Latin indices run from $0$ to $4$. We do not bother with Lagrange multipliers. For treating constraints we use the transformation (\ref{Trans6-4}) and pass to the coordinates that are perfectly adapted to the above constraints. In the cosmological constant limit the transformed Lagrangian takes the form
\begin{equation}
L_{\rm part}=\gamma^{-1}(\dot{z})\left(-m-\frac12\mu K_{4D}(\dot{z},\ddot{z})\right),
\end{equation}
where $\gamma(\dot{z})$ is the Lorentz factor and $K_{4D}(\dot{z},\ddot{z})$ is the four-dimensional counterpart of the curvature (\ref{k2}). In order to obtain the standard Lagrangian of a structureless point charge in four dimensions we should put $\mu=0$. However, we can interpret this renormalization constant as the desired manifestation of extra dimensions. Not only the constant, but also the term coupled with it. The latter yields the additional terms to the standard orbital momentum:
$$
M^{\alpha\beta}=z^\alpha P^\beta-z^\beta P^\alpha + q^\alpha\pi^\beta-q^\beta\pi^\alpha.
$$
After quantization, the squared curvature produces the spin one-half states \cite{PavAACA2015}. According to this paper, ``... the algebra of the Dirac brackets between the dynamical variables associated with velocity and acceleration contains the spin tensor'', namely
\begin{align}
s^{\alpha\beta}&=q^\alpha\pi^\beta-q^\beta\pi^\alpha\nonumber\\
&=\mu\left(a^\alpha u^\beta-a^\beta u^\alpha\right).\label{s2}
\end{align}
Let us evaluate the magnitude of constant $\mu$ for electron. The square of the spin tensor is $s^{\alpha\beta}s_{\alpha\beta}=-2\mu^2(a\cdot a):=-2\mu^2K$. According to Section \ref{RestFrame}, the squared curvature $K=(2m/\mu)k$ where dimensionless variable $k<1$. Equating the squared magnitude of the spin tensor to the squared electron's spin we get a rough estimate
$$
\mu_e\sim \frac{\hbar^2}{16m_e}\simeq 7.63\cdot 10^{-39}\mathrm{J}\mathrm{s}^2,
$$
where $\hbar$ is the Planck constant and $m_e$ is electron's mass in SI units. (We take into account that the squared acceleration contains the factor $c^2$.)

\section{Conclusions}\label{Concl}

Within the braneworld scenario, we consider dynamics of a classical charge in flat spacetime of six dimensions. Charge's electromagnetic field satisfies the Maxwell equations. A consistent regularization procedure which exploits the Poincar\'{e} symmetry of the theory results the particle action functional which contains, apart from usual ``bare'' mass, an additional renormalization constant coupled with the squared curvature of particle's world line \cite{KosTMP99,YarJPA04}. The mass shell of a free charge depends on the squared six-acceleration. This circumstance yields non-trivial inertial properties of rigid charge. The mass shell admits the time-like periodic orbit parameterized by the Jacobi elliptic functions.

We study the algebraic properties of electromagnetic field in 6D. We find three invariants of this field, establish their domain, and elaborate the classification scheme. We study the evolution of a charge acted upon a static homogeneous electromagnetic field. Privileged reference frames are defined for various types of field where it is simplified substantially. In this frame the world line is the combination of hyperbolic and circular orbits in three mutually orthogonal sheets of two dimensions.

Extra dimensions have been compactified by means of projection onto the four-dimensional de Sitter space embedded in the flat spacetime of six dimensions. We obtain the action integral point particle with rigidity/ The model is quantized in Ref.~\cite{PavAACA2015}. It is shown that the squared curvature leads to the spin states. Therefore, spins of elementary particles indicate presence of extra dimensions.

\begin{acknowledgments}
This research has been supported by Grant No. 0117U002093 of the National Academy of Science of Ukraine.
\end{acknowledgments}

\appendix

\section{Privileged reference frames}\label{PrivRF}

In this Appendix we present Lorentz matrices defining the inertial frames where the field strengths are functions of invariants (\ref{S-inv})-(\ref{P-inv}) only. We start with general case ${\cal P}\neq 0$ (see item (A) in the Classification list \ref{ClassF}). 
\begin{table*}
\caption{\label{EACC0sc}
Scalar products of five-vectors involved in the expression (\ref{Vect_eig}) for space components of eigenvector $\textbf{u}_\lambda$.}
\begin{ruledtabular}
\begin{tabular}{@{}l  c  c  c  c    c@{}}
& $\mathbf{E}$ & $\mathbf{B}$ & $\mathbf{A}$ & $\mathbf{C}$ & $\mathbf{C}_0$  \\
    \hline
$\mathbf{E}$  & $\mathbf{E}^2$ & ${\cal P}$ & $0$ & ${\cal Q}+\mathbf{B}^2$& $0$ \\
$\mathbf{B}$  & ${\cal P}$ & $\mathbf{B}^2$ & $0$ & ${\cal P}({\cal S}+\mathbf{E}^2)$ & $0$ \\
$\mathbf{A}$  & $0$ & $0$ &$\mathbf{E}^2({\cal S}+\mathbf{E}^2)-({\cal Q}+\mathbf{B}^2)$&$0$& $\mathbf{E}^2\mathbf{B}^2-{\cal P}^2$ \\
$\mathbf{C}$  & ${\cal Q}+\mathbf{B}^2$ & ${\cal P}({\cal S}+\mathbf{E}^2)$ & $0$ & ${\cal Q}\left({\cal S}+\mathbf{E}^2\right)+{\cal S}\mathbf{B}^2+{\cal P}^2$ & $0$ \\ 
$\mathbf{C}_0$& $0$ & $0$ & $\mathbf{E}^2\mathbf{B}^2-{\cal P}^2$ & $0$ & $({\cal Q}+\mathbf{B}^2)\mathbf{B}^2-({\cal S}+\mathbf{E}^2){\cal P}^2$
  \end{tabular}
\end{ruledtabular}
\end{table*}

\subsection{${\cal P}\neq 0$.}

Solving the matrix equation $\hat{F}\textbf{u}=\lambda\textbf{u}$ for a given eigenvalue $\lambda$ from the set (\ref{lbd_eig}) we derive the eigenvector $\textbf{u}_\lambda=(u_\lambda^0,u_\lambda^i)$ where zeroth component is completely arbitrary and the space components are
\begin{eqnarray}\label{Vect_eig}
u_\lambda^i&=&\frac{u_\lambda^0}{\lambda\left[\lambda^4+\left({\cal S}+\textbf{E}^2\right)\lambda^2+\textbf{B}^2\right]}\times\nonumber\\
&\times&\left(\lambda^4E^i+\lambda^3A^i+\lambda^2C^i+\lambda C_0^i+{\cal P}B^i\right).
\end{eqnarray}
Here $E^i$ and $B^i$ are the components of the electric field $\textbf{E}$ and magnetic field $\textbf{B}$ five-vectors, respectively. The other capital letters denote the components of five-vectors $\textbf{A}$, $\textbf{C}$, and $\textbf{C}_0$:
\begin{eqnarray}
A^i&=&\sum_{j\neq i}F_{ij}E^j,\nonumber\\
C^i&=&(-1)^{i+1}\sum_{j<k<l<r}\left(F_{j[k}G_{lr]}+G_{j[k}F_{lr]}\right),\nonumber\\
C^i_0&=&-\sum_{j\neq i}G_{ij}B^j.
\end{eqnarray}
where $(j,k,l,r)\neq i$ and square brackets in down indices denote circular permutation, e.g.
$C^1=F_{23}G_{45}+F_{25}G_{34}-F_{42}G_{53}+G_{23}F_{45}+G_{25}F_{34}-G_{42}F_{53}$.

We compose the matrix $U$ whose columns are the six eigenvectors (\ref{Vect_eig}) ranged according to the range of eigenvalues in the list (\ref{lbd_eig}). To derive the Lorentz matrix we multiply it on the auxiliary matrix $J$ given by eq.~(\ref{J}):
\begin{widetext}
\begin{equation}\label{Gar-B4}
(\Lambda^\mu{}_\nu)=\left(
\begin{array}{cccccc}
\frac{\displaystyle u^0_b-u^0_{-b}}{\displaystyle \sqrt{2}}&\frac{\displaystyle u^0_b+u^0_{-b}}{\displaystyle \sqrt{2}}&\frac{\displaystyle {\rm i}\left(u^0_a-u^0_{-a}\right)}{\displaystyle \sqrt{2}}&\frac{\displaystyle u^0_a+u^0_{-a}}{\displaystyle \sqrt{2}}&\frac{\displaystyle {\rm i}\left(u^0_c-u^0_{-c}\right)}{\displaystyle \sqrt{2}}&\frac{\displaystyle u^0_c+u^0_{-c}}{\displaystyle \sqrt{2}}\\
\frac{\displaystyle u^k_b-u^k_{-b}}{\displaystyle \sqrt{2}}&\frac{\displaystyle u^k_b+u^k_{-b}}{\displaystyle \sqrt{2}}&\frac{\displaystyle {\rm i}\left(u^k_a-u^k_{-a}\right)}{\displaystyle \sqrt{2}}&\frac{\displaystyle u^k_a+u^k_{-a}}{\displaystyle \sqrt{2}}&\frac{\displaystyle {\rm i}\left(u^k_c-u^k_{-c}\right)}{\displaystyle \sqrt{2}}&\frac{\displaystyle u^k_c+u^k_{-c}}{\displaystyle \sqrt{2}}
\end{array}
\right).
\end{equation}
\end{widetext}
We assume $u^0_{-b}=-u^0_b$, $u^0_{-a}=u^0_a$, and $u^0_{-c}=u^0_c$. Inserting these and the space components we obtain the following matrix elements:
\begin{align}
\Lambda^0{}_0&=\sqrt{2}u^0_b,\; \Lambda^i{}_0=\frac{\sqrt{2}u^0_b\left(b^2A^i+C^i_0\right)}{b^2\left(a^2+c^2+\mathbf{E}^2\right)+\mathbf{B}^2};\nonumber\\
\Lambda^0{}_1&=0,\; \Lambda^i{}_1=\frac{\sqrt{2}u^0_b\left(b^4E^i+b^2C^i+{\cal P}B^i\right)}{b\left[b^2\left(a^2+c^2+\mathbf{E}^2\right)+\mathbf{B}^2\right]};\label{Lbd0ib}\\
\Lambda^0{}_2&=0,\; \Lambda^i{}_2=\frac{\sqrt{2}u^0_a\left(-a^4E^i+a^2C^i-{\cal P}B^i\right)}{a\left[a^2\left(-b^2+c^2+\textbf{E}^2\right)-\textbf{B}^2\right]};\nonumber\\
\Lambda^0{}_3&=\sqrt{2}u^0_a,\; \Lambda^i{}_3=\frac{\sqrt{2}u^0_a\left(a^2A^i-C^i_0\right)}{a^2\left(-b^2+c^2+\textbf{E}^2\right)-\textbf{B}^2};\label{Lbd0ia}\\
\Lambda^0{}_4&=0,\; \Lambda^i{}_4=\frac{\sqrt{2}u^0_c\left(c^4E^i-c^2C^i+{\cal P}B^i\right)}{c\left[c^2\left(b^2-a^2-\mathbf{E}^2\right)+\mathbf{B}^2\right]};\nonumber\\
\Lambda^0{}_5&=\sqrt{2}u^0_c,\; \Lambda^i{}_5=\frac{\sqrt{2}u^0_c\left(-c^2A^i+C^i_0\right)}{c^2\left(b^2-a^2-\mathbf{E}^2\right)+\mathbf{B}^2}.\label{Lbd0ic}
\end{align}

To provide the general condition on the Lorentz transformation
\begin{equation}\label{eta-Lbd}
\eta=\Lambda^T\eta\Lambda.
\end{equation}
we fix the zeroth components of eigenvectors as follows:
\begin{eqnarray}
u^0_b\sqrt{2}&=&\sqrt{\frac{b^2\left(a^2+c^2+\textbf{E}^2\right)+\textbf{B}^2}{\left(b^2+a^2\right)\left(b^2+c^2\right)}},
\label{Lbd00b}\\
u^0_a\sqrt{2}&=&\sqrt{\frac{a^2\left(-b^2+c^2+\textbf{E}^2\right)-\textbf{B}^2}{\left(a^2-c^2\right)\left(a^2+b^2\right)}},
\label{Lbd00a}\\
u^0_c\sqrt{2}&=&\sqrt{\frac{c^2\left(b^2-a^2-\textbf{E}^2\right)+\textbf{B}^2}{\left(a^2-c^2\right)\left(c^2+b^2\right)}}.
\label{Lbd00c}
\end{eqnarray}
Direct calculations result the scalar products of five-vectors involved in the expression (\ref{Vect_eig}) for space part of eigenvector. They are presented in Table~\ref{EACC0sc}. Using these expressions one can easily verify that the matrix $\Lambda=UJ$ satisfy the equation (\ref{eta-Lbd}).

\subsection{${\cal P}=0$, ${\cal Q}\ne 0$.}


Let us consider the item (B) in the Classification list \ref{ClassF}. If the invariant ${\cal P}$ vanishes, the characteristic polynomial (\ref{eig_v}) simplifies:
\begin{equation}\label{eig_vP0}
\left(\lambda^4+{\cal S}\lambda^2-{\cal Q}\right)\lambda^2=0.
\end{equation}
The roots of fourth degree polynomial in the square brackets are
\begin{equation}
\lambda_\pm^2=\frac12\left(-{\cal S}\pm\sqrt{{\cal S}^2+4{\cal Q}}\right). \label{Rt-Q}
\end{equation}
Putting ${\cal P}=0$ in eq.~(\ref{Vect_eig}) and substituting $-{\cal S}\lambda^2+{\cal Q}$ for $\lambda^4$, we obtain
the space components of eigenvectors associated with the non-zero eigenvalues:
\begin{equation}\label{Vect_eigP0Qpl}
u_\lambda^i=\frac{u_\lambda^0}{\textbf{E}^2\lambda^2+{\cal Q}+\textbf{B}^2}\left(\lambda^3E^i+\lambda^2A^i+\lambda C^i+C_0^i\right).
\end{equation}
Zeroth components are completely arbitrary. The formula gives four columns of matrix $U$ which diagonalizes the electromagnetic field tensor.

The geometric multiplicity of zero eigenvalue is equal to its algebraic multiplicity. Jordan canonical form consists of two Jordan blocs composed from zeroes. Direct calculations result the following components of zero eigenvector
\begin{equation}\label{u00}
\mathbf{u}_0=\left(u_0^0,u_0^1,\left(G_{1i}u_0^0+B^iu_0^1\right)/B^1\right)^T,
\end{equation}
where $u_0^0$ and $u_0^1$ are completely arbitrary and index $i$ runs from $2$ to $5$. According to the Classification list \ref{ClassF}, there are two possibilities.

If ${\cal Q}>0$, the spectrum consists of double degenerate zero and two pairs, one real and the other pure imaginary (see eq.~(\ref{lbd_eigP0Qpl})). We design the auxiliary matrix as follows
\begin{equation}\label{JP0Qpl}
(J_{\alpha\beta})'=\frac{1}{\sqrt{2}}\left(
\begin{array}{cccccc}
1&1&0&0&0&0\\
-1&1&0&0&0&0\\
0&0&{\rm i}&1&0&0\\
0&0&-{\rm i}&1&0&0\\
0&0&0&0&\sqrt{2}&0\\
0&0&0&0&0&\sqrt{2}
\end{array}
\right).
\end{equation}
Putting $c=0$ in eqs.~(\ref{Lbd0ib}), (\ref{Lbd0ia}), (\ref{Lbd00b}), and (\ref{Lbd00a}) we obtain four columns of matrix $\Lambda=UJ'$ defining the transformation to privileged inertial frame. Keeping in mind eq.~(\ref{u00}) we assume the fifth column of matrix $U$ as
\begin{equation}\label{U-5}
u_0^0=0,\quad u_0^i=\frac{B^i}{\sqrt{\textbf{B}^2}}.
\end{equation}
The sixth column is
\begin{equation}\label{U-6}
u_0^0=\frac{\sqrt{\textbf{B}^2}}{ab},\quad u_0^i=\frac{C_0^i}{ab\sqrt{\textbf{B}^2}}.
\end{equation}
In the privileged inertial frame the electromagnetic field tensor is given by eq.~(\ref{FLP0Qpl}).

If $-{\cal S}^2/4<{\cal Q}<0$, both the squared eigenvalues (\ref{Rt-Q}) are negative. The spectrum consists of the double degenerate zero and two pairs of pure imaginary numbers (see eq.~(\ref{lbd-eigP0Qmn})). We define the auxiliary matrix
\begin{equation}\label{JP0Qmn}
(J_{\alpha\beta})''=\frac{1}{\sqrt{2}}\left(
\begin{array}{cccccc}
\sqrt{2}&0&0&0&0&0\\
0&\sqrt{2}&0&0&0&0\\
0&0&{\rm i}&1&0&0\\
0&0&-{\rm i}&1&0&0\\
0&0&0&0&{\rm i}&1\\
0&0&0&0&-{\rm i}&1
\end{array}
\right),
\end{equation}
The first column of the Lorentz matrix defining transformation to the privileged reference frame is
\begin{equation}\label{U-1}
u_0^0=\frac{\sqrt{\textbf{B}^2}}{ac},\quad u_0^i=\frac{C_0^i}{ac\sqrt{\textbf{B}^2}},
\end{equation}
while the second column is given by eq.~(\ref{U-5}). The others can be derived from the expressions (\ref{Lbd0ia}), (\ref{Lbd0ic}), (\ref{Lbd00a}), and (\ref{Lbd00c}) where $b=0$. In the privileged inertial frame the electromagnetic field tensor is given by eq.~(\ref{FLP0Qmn}).

\subsection{${\cal P}=0$, ${\cal Q}=0$, ${\cal S}\ne 0$.}

In this Paragraph we consider the item (C) of the Classification list \ref{ClassF}. We restrict ourselves to the non-defective zero eigenvalue when Jordan canonical form consists of four Jordan blocks composed from zeroes. Direct calculations produce the following components of zero eigenvectors:
\begin{equation}\label{u000}
\mathbf{u}_0=\left(u_0^i,\frac{E^5u_0^0+F_{5k}u_0^k}{F_{45}},
-\frac{E^4u_0^0+F_{4k}u_0^k}{F_{45}}\right)^T,
\end{equation}
where four components $u_0^i$, $i=0,1,2,3$, are completely arbitrary and index $k$ runs from $1$ to $3$.

As all the elements of Hodge dual matrix (\ref{Gar}) vanish, five-vectors $\mathbf{C}$ and $\mathbf{C}_0$ are identically equal to zero. Putting these and ${\cal Q}=0$ in eq.~(\ref{Vect_eigP0Qpl}) we obtain the space components of eigenvectors associated with the non-zero eigenvalues:
\begin{equation}\label{Vect_eigP0Q0pl}
u_\lambda^i=\frac{u_\lambda^0}{\textbf{E}^2}\left(\lambda E^i+A^i\right).
\end{equation}
Zeroth components $u_\lambda^0$ are completely arbitrary. The formula gives two columns of matrix $U$ which diagonalizes the electromagnetic field tensor. The scalar products are $(\mathbf{E}\mathbf{A})=0$ and $\mathbf{A}^2=\mathbf{E}^2\left({\cal S}+\mathbf{E}^2\right)$.

According to the Classification list \ref{ClassF}, there are two possibilities.

If ${\cal S}<0$, the spectrum consists of pair $+b$ and $-b$ where $b=\sqrt{-{\cal S}}$ and forth degenerate zero. We define the  auxiliary matrix in the form
\begin{equation}\label{JP0Q0Smn}
(J_{\alpha\beta})=\frac{1}{\sqrt{2}}\left(
\begin{array}{cccccc}
1&1&0&0&0&0\\
-1&1&0&0&0&0\\
0&0&\sqrt{2}&0&0&0\\
0&0&0&\sqrt{2}&0&0\\
0&0&0&0&\sqrt{2}&0\\
0&0&0&0&0&\sqrt{2}
\end{array}
\right).
\end{equation}
Simple calculations yield the first column of the Lorentz matrix $\Lambda=UJ$
\begin{equation}\label{Lbd_A}
\Lambda^0{}_0=\frac{|\mathbf{E}|}{b},\quad \Lambda^i{}_0=\frac{\mathbf{A}^i}{b|\mathbf{E}|},
\end{equation}
while the second column is
\begin{equation}\label{Lbd_E}
\Lambda^0{}_1=0,\quad \Lambda^i{}_1=\frac{\mathbf{E}^i}{|\mathbf{E}|}.
\end{equation}
The others should be composed from eigenvectors (\ref{u000}) by means of Gram-Schmidt process. In the privileged reference frame the electromagnetic field tensor is given by eq.~(\ref{FL-b}).

If the invariant ${\cal S}>0$, the Lorentz matrix is composed from four columns obtained with the help of Gram-Schmidt process, with column (\ref{Lbd_E}), and with modified column (\ref{Lbd_A}) where $b$ should be replaced by $a=\sqrt{{\cal S}}$. In the privileged reference frame the electromagnetic field tensor is given by eq.~(\ref{FL-a}).


%
%

%



\begin{thebibliography}{99}
\bibitem{S-TYN2010}
S-T. Yau and S. Nadis, {\it The Shape of Inner Space: String Theory and the Geometry of the Universe's Hidden Dimensions} (Basic Books, Reprint Ed., 2012).

\bibitem{CBR2016}
M. Chinaglia, A. E. Bernardini, and R. da Rocha, Int. J. Theor. Phys. {\bf 55}, 4605 (2016).

\bibitem{NA-HDD1998}
N. Arkani-Hamed, S. Dimopoulos, and G. Dvali, Phys. Lett. B {\bf 429}, 263 (1998).

\bibitem{AFK2015}
A. F. Ali, M. Faizal, and M. M. Khalil, Phys. Lett. B {\bf 743}, 295 (2015).

\bibitem{ATLAS2016}
ATLAS Collaboration, Phys. Lett. B {\bf 760}, 520 (2016).

\bibitem{C-BCL2006}
Y. Choquet-Bruhat, P.T. Chru\`sciel, and J. Loizelet, Class. Quantum Grav. {\bf 23}, 7383 (2006).

\bibitem{IvSk1951}
D. Ivanenko and A. Sokolov, {\it Classical Field Theory} (GosTehIzdat, Moskow, 2nd ed. (in Russian), 1951).

\bibitem{TY2012}
V. Tretyak and {\relax Yu} Yaremko, {\it Radiation Reaction in Classical Field Theory (Basics, Concepts, Methods)} (LAP Lambert Academic Publishing, Saarbr\"{u}ken, 2012).

\bibitem{KosIJMP2008}
B.P. Kosyakov, Int. J. Mod. Phys. A {\bf 23}, 4695 (2008).

\bibitem{Dirac1938}
P.A.M. Dirac, Proc. Roy. Soc. (London) A {\bf 167}, 148 (1938).

\bibitem{KosTMP99}
B.P. Kosyakov, Theor.\ Math.\ Phys. {\bf 119}, 493 (1999).

\bibitem{KLSprd02}
{\relax P.O. Kazinski}, {\relax S.L. Lyakhovich}, and {\relax A.A. Sharapov}, Phys. Rev. D {\bf 66}, 025017 (2002).

\bibitem{YarJPA04}
{\relax Yu} Yaremko, J.\ Phys. A: Math. Gen. {\bf 37}, 1079 (2004).

\bibitem{MMtmp2008}
{\relax A.D} Mironov and {\relax A.Yu} Morozov, Theor.\ Math.\ Phys. {\bf 156}, 1209 (2008).

\bibitem{PavAACA2015}
{\relax M. Pav\u{s}i\u{c}}, Adv. Appl. Clifford Algebras {\bf 26(1)}, 315 (2015).

\bibitem{PavPLB1988}
{\relax M. Pav\u{s}i\u{c}}, Phys. Lett. B {\bf 205(2-3)}, 231 (1988).

\bibitem{GILNPLB1989}
{{\relax J. Grundberg, J. Isberg, U. Lindstr{\relax\"{o}}m, and H. Nordstr{\relax\"{o}}m}}, Phys. Lett. B {\bf 231}, 61 (1989).

\bibitem{PolNPB1986}
A. Polyakov, Nucl. Phys. B {\bf 268(2)}, 406 (1986).

\bibitem{KlnPLB1986}
H. Kleinert, Phys. Lett. B {\bf 174(3)}, 335 (1986).

\bibitem{PlushPLB91}
M.S. Plyushchay, Phys. Lett. B {\bf 253}, 50 (1991).

\bibitem{KosBook07}
B.P. Kosyakov, {\it Introduction to the Classical Theory of Particles and Fields} (Springer, Heidelberg, 2007).

\bibitem{ASWactaB1989}
{\relax H. Arod{\relax\'{z}}, A. Sitarz, and P. W{\relax\c{e}}grzyn}, Acta Phys. Pol. B {\bf 20(11)}, 921 (1989).

\bibitem{DHOTPLB1990}
{\relax T. Dereli, D. H. Hartley, M. {\relax\"{O}}nder, and R. W. Tucker}, Phys. Lett. B {\bf 252(4)}, 601 (1990).

\bibitem{AbrStg}
M. Abramowitz and {\relax I.A. Stegun}, {\it Handbook of Mathematical Functions} (Dover Publications Inc., New York, 1964).

\bibitem{FradJPA78}
 D.M. Fradkin, J. Phys. A: Math. Gen. {\bf 11}, 1069 (1978).

\bibitem{YarJMP13}
{\relax Yu}. Yaremko, J. Math. Phys. {\bf 54}, 092901 (2013).

\bibitem{Schutz}
{\relax B.F.~Schutz}, {\it Geometrical methods of mathematical physics} (Campridge University Press, Campridge UK, 1980).

\bibitem{Aldro07}
R. Aldrovandi and {\relax J.P. Beltr{\relax \'{a}}n Almeida} and {\relax C.S.O. Mayor} and {\relax J.G. Perejra}, AIP Conf. Proc. {\bf 962}, 175 (2007).
\end{thebibliography}

\end{document}